\documentclass[12pt]{iopart}

\usepackage{iopams}  
\usepackage{graphicx}
\usepackage{caption}
 \usepackage{subfigure}
\begin{document}

\title[]{Statistical analysis of proton induced
reactions to generate recommended data for the production of medical radio-isotopes
}

\author{Sourav Mondal$^{1}$, A. Gandhi$^{1}$, and Rebecca Pachuau$^{1}$}

\address{$^1$Department of Physics, Banaras Hindu University, Varanasi 221005, India}
\ead{sourav131998@gmail.com-(Sourav Mondal)} \ead{pcrl.bec@gmail.com, rebecca24@bhu.ac.in (Corresponding author)-( Rebecca Pachuau)}
\vspace{10pt}
\begin{indented}
\item[]July 2022
\end{indented}

\begin{abstract}
Radio-isotopes produced via proton induced reaction holds special significance regarding nuclear medicine, astrophysical p-process, theragnostic and diagnostic processes. $^{76}$Br, $^{80m}$Br and $^{61}$Cu are positron emitter and they are useful in the functional studies via Positron Emission Tomography
(PET), whereas $^{77}$Br bears the potential for the application in Single Photon Emission Computed Tomography (SPECT) which involves electron capture process. PET and SPECT have been in high application in medical physics, diagnostics, therapy and nuclear medicine. $^{99m}$Tc and $^{64}$Cu are two popular radionuclide which play important role in nuclear medicine, currently being used in bio-medical physics, bone scan, modern imaging, blood pool leveling, oncology and diagnosis of copper related diseases. This paper focus on the generation of recommended nuclear reaction cross sections for the production of some useful medical radio-isotopes using the experimental datasets obtained from EXFOR database and simulated datasets from nuclear reaction model codes TALYS-1.95 and EMPIRE-3.1.1. 95\% confidence interval has been implemented to ensure confidence and precision. 
\end{abstract}
\maketitle
\section{\label{sec:level1}Introduction}
Study of cross section of proton induced reactions to produce medically significant different radio-isotopes have gained popularity in the last few decades enabling wide range of application in medical physics, diagnostic purpose, radiotherapy, nuclear medicine as well as in the detection and treatment of cell anomalies \cite{qaim2017nuclear}. In this article seven such radio-isotopes are considered stressing over their applications, suitable production route for production and generating statistically significant recommended cross sections which is the quantitative characteristics of nuclear reactions. Almost all the radionuclides are PET (Positron Emission Tomography) candidate which is an advance imaging system, having usage in diagnostic processes and analysis of the location of malfunction of human organs. $^{76}$Br is suitable for PET scan whereas $^{77}$Br is for SPECT (Single Photon Emission Computed Tomography).  $^{77}$Br and $^{80m}$Br are the two auger electron emitting radionuclides of bromine and are promising candidates for internal radiotherapy. Usage of bromine for radiopharmaceutical synthesis holds some advantages due to its high chemical versatility (including high nucleophilicity) enabling also electrophilic substitution making it suitable proliferation marker in case of local DNA synthesis rate \cite{breunig2015production}.

%%%%%%%%%%%%%%%%%%%%%%%%%%%%%%%%%%%%%%%%%%%%%%%%%%%%%%%%%%%%%%%%%%%%%%%%%%%%%%%%%%%%%%%%%%%%%%%%%%%%%%%%%%%%%%%%%%
The $^{100}$Mo(p,2n)$^{99m}$Tc reaction is one of the proposed processes for direct production of $^{99m}$Tc. The relative importance of the possible contributing reactions and their respective contribution to the total activity and the patient dose can be theoretically estimated and directly applicable in medical diagnostics. $^{99m}$Tc radionuclide is also widely used in SPECT, bone-scan, sentinel-node identification, cardiac ventriculography, functional brain imaging, myocardial perfusion imaging, immunoscintigraphy and blood pool labeling \cite{papagiannopoulou2017technetium}.

%%%%%%%%%%%%%%%%%%%%%%%%%%%%%%%%%%%%%%%%%%%%%%%%%%%%%%%%%%%%%%%%%%%%%%%%%%%%%%%%%%%%%%%%%%%%%%%%%%%%%%%%%%%%%%%%%%%%%%%%%%%%%%%%%%%%%%%%%%%%%%%%%%%%%%%%%%%%%%%%%%%%%%%%%%%%%%%%%%%%%%%%%%%%%%%%%%%%%%%%%%%%%%%%%%%%%%%%%%%%%%%%%%%%
The short-lived radionuclide $^{61}$Cu (T$_{1/2}$ = 3.3 hr) which is well suitable for PET imaging can be produced by the different contributing reactions. There is a growing interest in the use of $^{61}$Cu for diagnostic purposes for targeted radiotherapy. Here two different channels are discussed for the production of $^{61}$Cu and the decay data of the reactions along with the production routes and Q-value are tabulated in Table 1.

%%%%%%%%%%%%%%%%%%%%%%%%%%%%%%%%%%%%%%%%%%%%%%%%%%%%%%%%%%%%%%%%%%%%%%%%%%%%%%%%%%%%%
The relatively low threshold energy for the $^{64}$Ni(p,n) reaction has direct indication that $^{64}$Cu could be produced in therapeutic quantities even with low energy accelerators. $^{64}$Cu has a longer half-life than most positron-emitters (12.7 hr) and is thus ideal for diagnostic PET imaging of bio molecules. $^{64}$Cu is the only known radionuclide in the nuclear medicine transmuting through three different routes namely positron decay, beta decay and electron capture. Consequently, these result in emission of positron particles which can be used in (PET) and $\beta$ particles and auger electrons both of which find use in therapy. Thus, this unique nature allows $^{64}$Cu to be used for both diagnostic and therapeutic purpose in nuclear medicine which has enhanced its popularity. $^{64}$Cu is linked to a variety of bio molecules and can be used as theragnostic agents in human malignancies such as prostate, breast cancers, melanoma, glioblastoma and also diagnosis of human copper-associated diseases such as atherosclerosis, alzheimer’s disease etc.
%%%%%%%%%%%%%%%%%%%%%%%%%%%%%%%%%%%%%%%%%%%%%%%%%%%%%%%%%%%%%%%%%%%%%%%%%%%%%%%%%%%%%%%%%%%%%%%%%%%%%%%%%%%%%%%%%%%%%%%%%%%%%%%%%%%%%%%%%%%%%%%%%%%%%%%%%%%%%%%%%%%%%%%%%%%%%%
In the field of radionuclide production for medical applications, it is sophisticated way to accompany the experimental cross section with the theoretical calculation of the same, using a suitably developed simulation software, based on the theory of the nuclear reaction models involved \cite{csekerci2020level,ozdougan2021estimations,yiugit2014nuclear}.
Besides being of interest for the development of nuclear models, the measurement of nuclear reaction cross sections over a broad energy range is also a crucial step in the determination of optimum conditions for a production route. The cross sections from statistical models are sensitive to various nuclear physics input parameters, like nuclear level densities, optical potentials, the gamma-ray strength functions etc. For one particular reaction one can see variation of calculated parameters depending on the level density parameter and gamma ray strength function parameters. All sensitivities are considered in the calculations here and optimum path is selected hence. Theoretical calculation is performed and hence comments has been made whether theoretical models go with experimental outcome or there is partial disagreement between them \cite{bakhtiari2021production}. The latest decay data of radio-isotopes and suitable production channel of all the reactions studied in the present work are presented in Table 1.
%%%%%%%%%%%%%%%%%%%%%%%%%%%%%%%%%%%%%%%%%%%%%%%%%%%%%%%%%%%%%%%%%%%%%%%%%%%%%%%%%%%%%%%%%%%%%%%%%%%%%%%%%%%%%%%%%%%%%%%%%%%%%%%%%%%%%%%%%%%%%%%%%%%%%%%%%%%%%%%%%%%%%%%%%%%%%%
\section{\label{sec:level2}Nuclear Model Code Calculation}
\subsection{TALYS-1.95 Nuclear Reaction Model Code}
In TALYS-1.95 nuclear model code simulation, one can simulate the nuclear reactions that involves particles as projectiles in the wide energy range from 1 keV to 200 MeV~\cite{koning2012modern}. The code consists of 6 different nuclear level density models (ldmodel 1-6)  providing a wide range of opportunity for simulating nuclear reactions \cite{pachuau2021fast,pachuau2019neutron}. The basic objective behind its construction is the simulation of nuclear reactions that involve neutrons, photons, protons, deuterons, tritons, $^{3}$He and alpha-particles as projectiles and for target nuclides of mass 12 and heavier. The nuclear model calculations based on TALYS-1.95 is generally categorized into Direct, Pre-equilibrium and Compound nuclear reactions and specific statistical theories have been implemented in it \cite{gandhi2019cross,aydin2015comparison,xu2014systematic,fuladvand2013pre,goriely2008improved}.

The default optical model parameters (OMP) used in TALYS-1.95 are based on local and
global parametrizations by Koning and Delaroche  \cite{koning2003local}. A brief overview about
level density models in TALYS-1.95 is given below
\begin{enumerate}
\item  ldmodel-1: The constant temperature and Fermi-gas model
where the constant temperature model is used in the low excitation region and the Fermi-gas model is used in the high excitation energy region. The transition
energy is around the neutron separation energy.
\item ldmodel-2: The back shifted Fermi-gas model.
\item ldmodel-3: The generalized superfluid model.
\item ldmodel-4: The microscopic level densities (Skyrme force) from Goriely's tables.
\item ldmodel-5: The microscopic level densities (Skyrme force) from Hilaire’s combinatorial tables \cite{zhao2020microscopic}.
\item ldmodel-6: The microscopic level densities (temperature-dependent Hartree-Fock Bolyubov, Gogny force) from Hilaire’s combinatorial
tables~\cite{hilaire2012temperature}. \\
\end{enumerate}

In statistical models for predicting cross sections, nuclear level densities are used
at excitation energies where discrete level information is not available or
incomplete. We use several models for the level density in TALYS-1.95, which ranges from
phenomenological analytical expressions to tabulate level densities derived from
microscopic models.

Fermi gas model: the best-known analytical level density model to explain the
density of states is Fermi Gas Model (FGM). Here it is assumed that collective levels
in the deformed nucleus are absent and equ-spaced excited states of single particle
have been taken into consideration. In this model Fermi gas state density are
dependent on level density parameter, spin cut off parameter which represents the
width of the angular momentum distribution and energy shift which is an empirical
parameter.

Constant Temperature Model: In this model the excitation energy range is divided
into a low energy part where constant temperature law holds and a high energy part
above where the Fermi gas model applies. The matching energy expressions are
dependent on the mass number of the nuclei and energy shift.

The Back shifted Fermi gas model: In this model the pairing energy is treated as an
adjustable parameter and here also Fermi gas expression is used all the way down
to 0 MeV. Here energy shift due to Back shifted Fermi gas model is dependent on
pairing energy parameter, mass number of the nuclei and also an extra adjustable
parameter to fit the experimental data per nucleus. A slight problem regarding
diverging solution as critical energy U tends to zero has been considered in this model and also
additional spin distributions are also included.

Generalized Superfluid Model and Microscopic level densities: Based on the Bardeen-Cooper-Schrieffer theory
superconductive pairing correlation is taken into account in this theory. This model
is analyzed by the phase transition from a superfluid behavior at low energy. Where
pairing correlation s strongly influence the level density. Here the critical level
density parameter is given by iterative equation with known parameters. Critical
entropy, critical determinant and spin cut-off parameter are also included.

Microscopic Level Densities: By the calculations of S. Goriely based on Hartree-Fock
Approximation different level densities are have been calculated up to energy
150~MeV. Moreover, new energy, spin and parity-dependent nuclear level densities
based on the microscopic combinatorial model have been proposed by Hilaire and
Goriely. The only phenomenological aspect of the model is a simple damping
function for the transition from spherical to deformed. The calculations make
coherent use of nuclear structure properties determined within the deformed
Skyrme-Hartree-Fock-Bogolyubov framework.\\
There are eight photon strength functions available in TALYS-1.95 code which are discussed below-
(i) Photon strength function 1 (PSF-1): Kopecky-Uhl
generalized Lorentzian \cite{kopecky1990test}.\\
(ii) Photon strength function 2 (PSF-2): Brink \cite{brink1957individual} and
Axel Lorentzian \cite{axel1962electric}.\\
(iii) Photon strength function 3 (PSF-3): Hartree-Fock
BCS tables \cite{capote2009ripl}.\\
(iv) Photon strength function 4 (PSF-4): Hartree-FockBogolyubov tables \cite{capote2009ripl}.
In our present work we considered different photon strength functions in the definite ldmodels to ensure best results comes out.\\
(v) Photon strength function 5 (PSF-5): Goriely’s hybrid
model \cite{goriely1998radiative}.\\
(vi) Photon strength function 6 (PSF-6): Goriely
temperature-dependent Hartree-Fock-Bogolyubov.\\
(vii) Photon strength function 7 (PSF-7): Temperature dependent relativistic mean field.\\
(viii) Photon strength function 8 (PSF-8): Gogny D1M
Hartree-Fock-Bogolyubov+QRPA.\\
For each level density model we can incorporate eight gamma ray strength function based on which we can have wide range of variation for simulating nuclear reactions \cite{pachuau2018energy,ozdougan2019investigation,larsen2010impact}.
%%%%%%%%%%%%%%%%%%%%%%%%%%%%%%%%%%%%%%%%%%%%%%%%%%%%%%%%%%%%%%%%%%%%%%%%%%%%%%%%%%%%%%%%%%%%%%%%%%%%%%%%%%%%%%%%%%%%%%%%%%%%%%%%%%%%%%%%%%%%%%%%%%%%%%%%%%%%%%%%%%%%%%%%%%%%%%
\subsection{EMPIRE-3.1.1 Nuclear Reaction Model Code}
EMPIRE-3.1.1 code is adaptable and can perform whole nuclear reaction cross section calculations covering a vast energy range up to several MeV. In this code four level density models (LEVDEN-0,1,2,3) are available for simulation of nuclear reaction cross sections. The physics core of EMPIRE-3.1.1 is a fusion of the original code and codes written by other authors to match theoretical prediction with experimental outcomes~\cite{herman2007empire}.
%The following codes are incorporated in the physics core of the EMPIRE:\\
%\textbf{ECIS03} Coupled-Channels and DWBA code by J. Raynal\\
%\textbf{CCFUS} simplified Coupled-Channels calculation of
%HI fusion cross section by C.H. Dasso and S.
%Landowne\\
%\textbf{ORION \& TRISTAN TUL}  approach to multistep direct by H. Lenske \\
%\textbf{DEGAS} exciton model with angular momentum conservation and gamma emission  by E. B-etak and P.Oblo-zinsky,\\
%\textbf{DDHMS} Monte-Carlo simulation of the pre-equilibrium decay by M.B. Chadwick\\
%\textbf{BARMOM} fission barriers and moments of inertia by A. Sierk \\
In EMPIRE-3.1.1, the pre-equilibrium contribution in nuclear reaction cross-sections have been calculated using the phenomenological exciton model (via PCROSS code) that depends on the particle-hole state level densities. For the compound nuclear reaction cross-sections, the statistical model based on Hauser–Feshbach formalism has been used. The optical model parameters used in the calculations are taken from the RIPL-3 (Reference Input Parameter Library), in which the parameters proposed by Koning and Delaroche have been used for the outgoing protons and neutrons, whereas the parameters of Avrigeanu et al. \cite{Avri} have been used for the outgoing $\alpha$ particles. The gamma-ray strength function (GSTRFN-1) described by the modified Lorentzian (MLO1) has been used in the compound nucleus model calculations of particle emission~\cite{gst}. The parameters used in the model calculation have been retrieved from the RIPL-3~\cite{capote2009ripl}. 

For the level densities, the default phenomenological level density (LEVDEN-0) is Enhanced Generalized Superfluid Model (EGSM) which is the default model in the EMPIRE calculation. In EGSM, below the critical energy U, the nuclear level density is calculated according to GSM, the superfluid model and according to FGM above critical energy. Whereas LEVDEN-1,2,3 are described by Generalized Superfluid Model, Gilbert-Cameron Level densities and RIPL-3 microscopic HFB level densities respectively. Up to LEVDEN-2 the  models are phenomenological in nature but LEVDEN-3 is microscopic combinatorial approach. The details of the models used for the calculations in this work are presented in Table 2.
%%%%%%%%%%%%%%%%%%%%%%%%%%%%%%%%%%%%%%%%%%%%%%%%%%%%%%%%%%%%%%%%%%%%%%%%%%%%%%%%%%%%%%%%%%%%%%%%%%
\begin{table*}
\caption{Decay data and production route of product radio-nuclides
}
\begin{scriptsize}
\begin{tabular}{|c|c|c|c|c|c|c|}
\hline
\hline
Nuclei & Half-life (hr) & Decay Mode & Gamma Intensity (\%) & Energy (keV) & Production Route & Q value (MeV)\\
\hline
\hline
$^{76}$Br & 16.2 & EC: 45\% & 74.0 & 559.09&$^{76}$Se(p,n)$^{76}$Br& 5.822\\
&&$\beta^{+}$: 55\%&&&&\\
\hline
$^{77}$Br & 57.03 & EC: 100\% & 23.1 & 238.9 &$^{77}$Se(p,n)$^{77}$Br& 2.175\\

\hline
$^{80m}$Br &4.4&IT = 100\%&39.1&37.05&$^{80}$Se(p,n)$^{80m}$Br& -2.652\\
\hline
$^{61}$Cu&3.4&$\beta^{+}$: 61\%&12.5&283&$^{61}$Ni(p,n)$^{61}$Cu & -3.02\\
 &3.33&&12.2 & 283&$^{64}$Zn(p,$\alpha$)$^{61}$Cu&0.844\\
\hline
$^{64}$Cu &12.7&$\beta^{+}$: 18\%&36&511&$^{64}$Ni(p,n)$^{64}Cu$& -2.46\\
\hline
$^{99m}$Tc &6.0082&IT: 99.9963\%&89.5&140.5&$^{100}$Mo(p,2n)$^{99m}$Tc & -7.72\\
\hline
\end{tabular}

\end{scriptsize}\end{table*}
%%%%%%%%%%%%%%%%%%%%%%%%%%%%%%%%%%%%%%%%%%%%%%%%%%%%%%
%%%%%%%%%%%%%%%%%%%%%%%%%%%%%%%%%%%%%%%%%%%%%%%%
\section{\label{sec:level1}Evaluation procedure and Statistical data analysis}
From the knowledge of statistical analysis, it can be obtained that fitted parameters for the goodness of the fit test are chi-square, r-squared and adjusted r-squared. In this work,  95\% confidence limit is chosen and calculated using Origin software for all the reactions. As our theoretical uncertainty is given by 95\% confidence interval, the uncertainty width depends completely on the data distribution. The better the data distribution, the shorter is the confidence interval.

In the initial calculations, the theoretical and the experimental cross sections do not agree in a good manner. Hence, 3$\sigma$ range had been calculated for all the reactions using both TALYS-1.95 and EMPIRE-3.1.1. However the 3$\sigma$  range calculated using default parameters of TALYS-1.95 do not match with that of EMPIRE-3.1.1. 
%Problems arise for the calculation of recommended energy range as the theoretical calculation for the codes and experimental observation do not match in a straight forward manner. Separately, 3-sigma ranges for all the reactions have been calculated for both the reaction model code and there is a problem faced as 3-sigma ranges do not match. Hence, different level density models can be implemented to ensure that the measured cross section by theoretically calculated cross section does not fluctuate so much. 
In order to obtain the best agreement, different level density models and photon strength functions of TALYS-1.95 code have been explored. Although for some reactions, the experimental cross sections are consistent to each other, others are quite scattering in nature. That is why, in addition to the level density models the photon strength functions are also varied to obtain the best recommended data using the TALYS-1.95 and EMPIRE-3.1.1. 3$\sigma$ range is calculated from the measured by theoretically calculated cross sections and hence the data points lying outside 3$\sigma$  are excluded and then fitting is implemented. Orthogonal distance regression algorithm and polynomial fit curve generates best results for fitting.

Polynomial curve is fitted first through all the measured by calculated data points thereafter and best fitted ratio factor is generated. No weighting with fixed intercept 10 is considered for best fitting. From the ideal best fit curve (r-squared $>$ 0.99), best fit ratio is obtained for those particular energies from both the nuclear reaction model codes for each reaction and recommended data is produced via normalisation. The recommended data are very smooth in nature and have low uncertainties. 95\% confidence interval is furthermore associated with the recommended data and we see very minute variation or uncertainty. Here the estimation of recommended data and confidence interval are completely theoretical approximation. The recommended cross sections at different proton energies where no measurements had been done are also generated and presented in Table 3 and Table 4.
%%%%%%%%%%%%%%%%%%%%%%%%%%%%%%%%%%%%%%%%%%%%%%%%%%%%%%%%%%%%%%%%%%%%%%%%%%%%%%%%%%
\begin{table*}
\caption{Statistical parameters for comparison of experimental cross sections and theoretical predictions.
}
\begin{scriptsize}
\begin{tabular}{|c|c|c|c|}
\hline
\hline
Proton Induced Reactions&  ldmodel/LEVDEN and PSF/GSTRFN&TALYS-1.95 parameters& EMPIRE-3.1.1 parameters\\
\hline
$^{76}Se(p,n)^{76}$Br& & D=0.3168 & D= 0.2877\\

&TALYS-1.95: ldmodel-3 and PSF-2&F=0.3592 & F=0.3829 \\
&&R=0.8603 & R= 1.0706\\
&EMPIRE-3.1.1: LEVDEN-3 and GSTRFN-1&K=1.0592 & K=1.0452 \\
\hline
\hline
$^{77}Se(p,n)^{77}$Br & &D=0.2964 & D=0.4770 \\

&TALYS-1.95: ldmodel-3 and PSF-2&F=0.4949 & F=0.5442 \\
&&R=1.1417 & R=1.1773\\
&EMPIRE-3.1.1: LEVDEN-3 and GSTRFN-1&K=1.0625 & K=1.4094 \\
\hline
\hline
$^{80}Se(p,n)^{80m}$Br & &D=0.3218 & D=0.2846\\

&TALYS-1.95: ldmodel-1 and default PSF-1&F=0.3367 & F=0.3457 \\
&&R= 0.6781& R=0.8604 \\
&EMPIRE-3.1.1: LEVDEN-3 and GSTRFN-1&K=1.0742 & K= 1.4206\\
\hline
\hline
$^{61}Ni(p,n)^{61}$Cu && D=0.1927 & D=0.1919 \\

&TALYS-1.95: ldmodel-3 and default PSF-1&F=0.5012& F= 0.4817\\
&&R=1.0323 & R= 1.1300\\
&EMPIRE-3.1.1: LEVDEN-3 and GSTRFN-1&K=1.0565 & K=1.4589 \\
\hline
\hline
$^{64}$Zn(p,$\alpha$)$^{61}$Cu && D=0.3172 & D=0.1690 \\

&TALYS-1.95: ldmodel-2 and PSF-1&F=0.5107 & F=0.4116 \\
&&R=1.138 & R=1.1057 \\
&EMPIRE-3.1.1: LEVDEN-3 and GSTRFN-1&K=1.0456 & K= 1.0306\\
\hline
\hline
$^{100}$Mo(p,2n)$^{99m}$Tc && D=0.2869 & D=0.4086 \\
&TALYS-1.95: ldmodel-4 and PSF-2&F=0.3049 & F=0.2804 \\
&&R=0.7130 & R=1.03\\
&EMPIRE-3.1.1: LEVDEN-3 and GSTRFN-1&K=1.066 & K=1.3674\\
\hline
\hline
$^{64}$Ni(p,n)$^{64}$Cu && D=0.3172 & D=0.1680 \\
&TALYS-1.95: ldmodel-4 and default PSF-1&F=0.5107 & F=0.4116\\
&&R=1.138 & R=1.1057 \\
&EMPIRE-3.1.1: LEVDEN-3 and GSTRFN-1&K=1.0456& K= 1.0306\\
\hline
\end{tabular}

\end{scriptsize}\end{table*}
%%%%%%%%%%%%%%%%%%%%%%%%%%%%%%%%%%%%%%%%%%%%%%%%%%%%%%%%%%%%%%%%%%%%%%%%%%%%%%%%%%%%%
95\% confidence level is the factor which determines the upper and lower confidence boundary such that the probability of each measured by calculated cross section falling within this boundary is 0.95. If one increases the amount of confidence interval, one has to lose precision as fluctuating data must also be considered to ensure greater confidence and vice-versa. We went for 95\% as it will exclude those bad data ensuring great precision though having slightly lower precision. %Also 3$\sigma$ depends on data distribution.

3$\sigma$ is the statistical measure where 99.9996\% of data falls within this range thus we can neglect very few of experimental data. That is why a particular energy range can be chosen so that we can neglect highly fluctuating points as long as energy range consists of peak function data. Statistical parameters for each reaction has been  calculated and presented in Table 2. As the statistical parameter calculations for both TALYS-1.95 and EMPIRE-3.1.1 matches in a great manner, it can be said that it is significant to choose a particular ldmodel for TALYS-1.95 from the comparison curve in order to have consistent result with that of EMPIRE-3.1.1. It also signifies that deviation between theoretical prediction and experimental outcomes is almost same for both codes. If the data are scattering in nature, in that case, variance will have wide range so those bad data points have to be considered but it will lie outside of 95\% confidence as it indicates that including those data will lower the precision.\\

 For the choice of one particular ldmodel and one particular photon strength function to predict recommended energy range is well supported by these statistical parameters as these parameters are the direct measurement of fluctuations between theory and experiment \cite{takacs2003validation,saito2020production,mohr2012recommended,liu2021excitation}.  So, picking one ldmodel and one photon strength function from the rest to predict the 3$\sigma$ range from coinciding range of statistical parameters are evident. First the definitions of these statistical parameters are mentioned and thus the values of these parameters are presented in Table 2.

\begin{equation}
D (mean\;relative\;deviation)=\frac{1}{N}\sum_{i=1}^{N}\frac{\sigma_i^{cal}-\sigma_i^{exp}}{\sigma_i^{exp}}\\
\end{equation}

\begin{equation}
F (mean\;standerdised\;deviation)=\left\{\frac{1}{N}\sum_{i=1}^{N}\frac{ \sigma_i^{cal}-\sigma_i^{exp}}{\sigma_i^{exp}}\right\}^\frac{1}{2}\\
\end{equation}

\begin{equation}
R (mean\;ratio)=\frac{1}{N}\sum_{i=0}^{N} \frac{\sigma_i^{cal}}{\sigma_i^{exp}}
\end{equation}

\begin{equation}
K (mean\;square\;logarithmic\;deviation)=10^{\frac{1}{N}\sqrt{\sum_{i=0}^{N}\left[\log(\sigma_i^{exp})-\log(\sigma_i^{cal})\right]^2}}
\end{equation}
where ${\sigma}^{exp}$ and ${\delta\sigma}^{exp}$ are the experimental cross section and its uncertainty in the energy E and N is the number of data points in the measured cross-section. D is also called the relative variance, the smallest 
value of D, represents us the best quality of the overlapping between the model calculations and experimental data.
The relative standard deviation of a set of data can be depicted as either a percentage or as a number. The higher the relative standard deviation, the more spread out the results are from the mean of the data. On the other hand, a lower relative standard deviation means that the measurement of data is more precise. Mean ratio is the measure of fluctuation of data from one code to another which is the indirect indication of measured/calculated cross section used later. Mean squared logarithmic error (MSLE) can be interpreted as a measure of the ratio between the true and predicted values. Mean squared logarithmic error is, as the name suggests, a variation of the Mean Squared Error. MSLE only care about the percentual difference \cite{michalet2010mean,zdeb2013half,am2022}.
%%%%%%%%%%%%%%%%%%%%%%%%%%%%%%%%%%%%%%%%%%%%%%%%%%%%%%%%%%%%%%%%%%%%%%%%%%%%%%%%%%%%%%%%%%%%%%%%%%%%%%%%%%%%%%%%%%%%%%%%%%%%%%%%%%%%%%%%%%%%%%%%%%%%%%%%%%%%%%%%%%%%%%%%%%%%%%

%%%%%%%%%%%%%%%%%%%%%%%%%%%%%%%%%%%%%%%%%%%%%%%%%%%%%%%%%%%%%%%%%%%%%%%%%%

%%%%%%%%%%%%%%%%%%%%%%%%%%%%%%%%%%%%%%%%%%%%%%%%%%%%%%%%%%%%%%%%%%%%%%%%%%%%%%
\subsection{\label{sec:level2}Normalisation of cross section}
Normalization of cross section is done on the basis of latest decay data available on the NuDat-2.8 database \cite{ilic2020prompt}. This process must be done as it is the standard platform where one author’s analysis is easily compared with others. Here four parameters are used for normalization i.e., half-life, isotopic abundance, monitor cross-section and gamma-intensity. Normalisation is done on the basis of latest decay data.

\subsection{\label{sec:level2}95\% confidence width calculation}
By general definition of 95\% confidence interval one can interpret that errors from mean value is calculated from the variance of the data distribution. For a particular energy one measures cross section but depending on the accuracy of measurement this measured value differs. As we have small range of cross sections for each energy, the range is bounded by upper and lower boundary limit. 95\% confidence limit ensures that the probability of all such cross sections being measured within this range is 0.95  \cite{aslam2009charged,sudar2002measurements}. The error function is calculated from the critical parameter. So one can have theoretical uncertainty as well as experimental uncertainty for estimation.
Generally a parameter $\alpha$ is defined so that $\alpha=1-CL$, where $CL$=Confidence Limit. For 95\% confidence $\alpha=1-0.95 =0.05$\\
So, confidence interval =$\overline{x}\pm E$,\\
\hspace*{0.5cm} where $\overline{x}$=mean\\
\hspace*{1.7cm}$E$=error\\
We get error function expression $E=Z_{\frac{\alpha}{2}} \frac{\sigma}{\sqrt{n}}$\\
where $Z_{\frac{\alpha}{2}}$= critical parameter determined by the Origin Software\\
So, we get confidence interval =$Z_{\frac{0.05}{2}} \frac{\sigma}{\sqrt{n}}$\\
\hspace*{5.5cm}=$Z_{0.025}\frac{\sigma}{\sqrt{n}}$\\
where $\sigma$=standard deviation\\
\hspace*{1cm} $n$=number of data plots
\section{Results and Discussion}
\subsection{Measured by calculated cross section using TALYS-1.95 and EMPIRE-3.1.1 and recommended data evaluation}
As it is completely theoretical work, the cross sections of each reaction had been calculated thoroughly and then from the best fit ratio factor, measured by calculated cross section are plotted and the required recommended data are produced. In order to obtain the ratio of the measured cross sections available in the EXFOR database and the theoretically calculated cross sections for radionuclides produced by proton induced reactions on Se, Ni, Zn and Mo targets, theoretical simulations had been performed using TALYS-1.95 and EMPIRE-3.1.1.
The calculations are performed using TALYS-1.95 default ldmodels and PSF at first and then varying the ldmodels to study how the choice of different level density models can affect the results and also default EMPIRE-3.1.1 code predictions. For these reactions we plot the measured by calculated data along with the energy to see the fluctuations of experimental data with theoretical data. The goodness of the fit test is given by the r-squared value.

For fit the more r-squared value goes towards 1, better the fit is. For most of the reactions the reported experimental data available in the EXFOR database are scattered whereas theoretical simulated data curves are smooth in nature. In the higher energy region, the cross sections calculated using TALYS-1.95 best-y produces much lower values as compared to measured cross sections and EMPIRE-3.1.1 for all the reactions except $^{80}$Se(p,n)$^{80m}$Br reaction. So, it is good to go for each level density model and see for which level density model TALYS-1.95 recommended data and EMPIRE-3.1.1 recommended data range agrees well. To tackle the problem,  calculations were done with different level density models of TALYS-1.95 to see which ldmodel goes well with EMPIRE-3.1.1 and experimental data. It is a general motivation to minimize the value of measured by calculated cross sections ratio incorporating the fitting parameters so that theoretical model agrees with experimental measurements. After selecting a particular ldmodel, PSFs are then considered and one unique PSF is chosen on the basis of common $3\sigma$. Polynomial fit curve is fitted through measured by calculated cross section points and lastly best fit ratio factor is generated. The recommended cross sections are calculated using average cross section of two different codes after the consideration of best fit ratio factor. Furthermore, 95\% confidence interval is associated to involve uncertainty but these uncertainties are very small which means that our recommended cross sections have very little fluctuation. Finally, our recommended cross sections are compared with the existing experimental data.
%%%%%%%%%%%%%%%%%%%%%%%%%%%%%%%%%%%%%%%%%%%%%%%%%%%%%%%%%%%%%%%%%%%%%%
\begin{figure*}%
\begin{center}
\subfigure[][]{%
\label{fig:ex3-a}%
\includegraphics[height=2.3in]{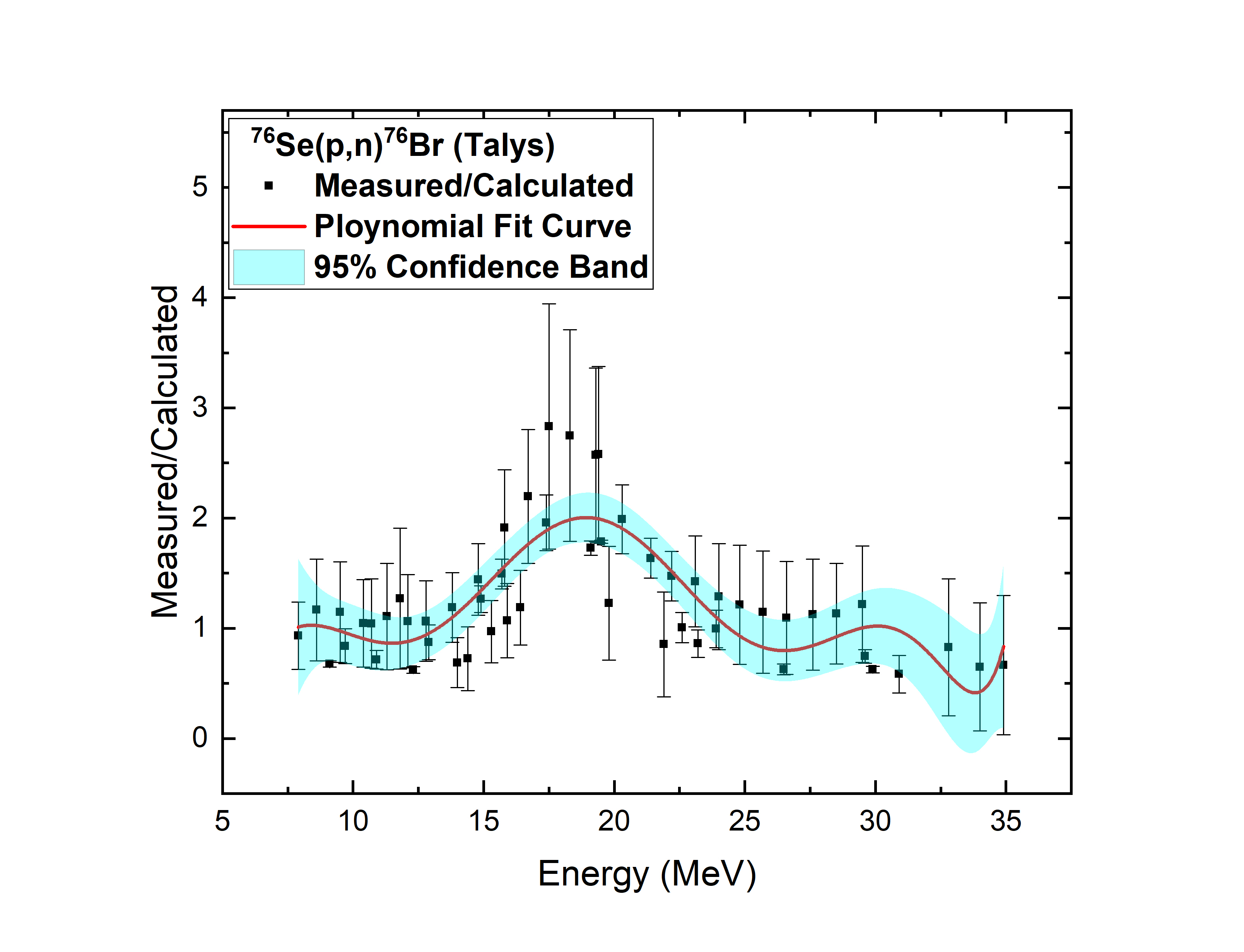}}%
\hspace{1pt}%
\subfigure[][]{%
\label{fig:ex3-b}%
\includegraphics[height=2.3in]{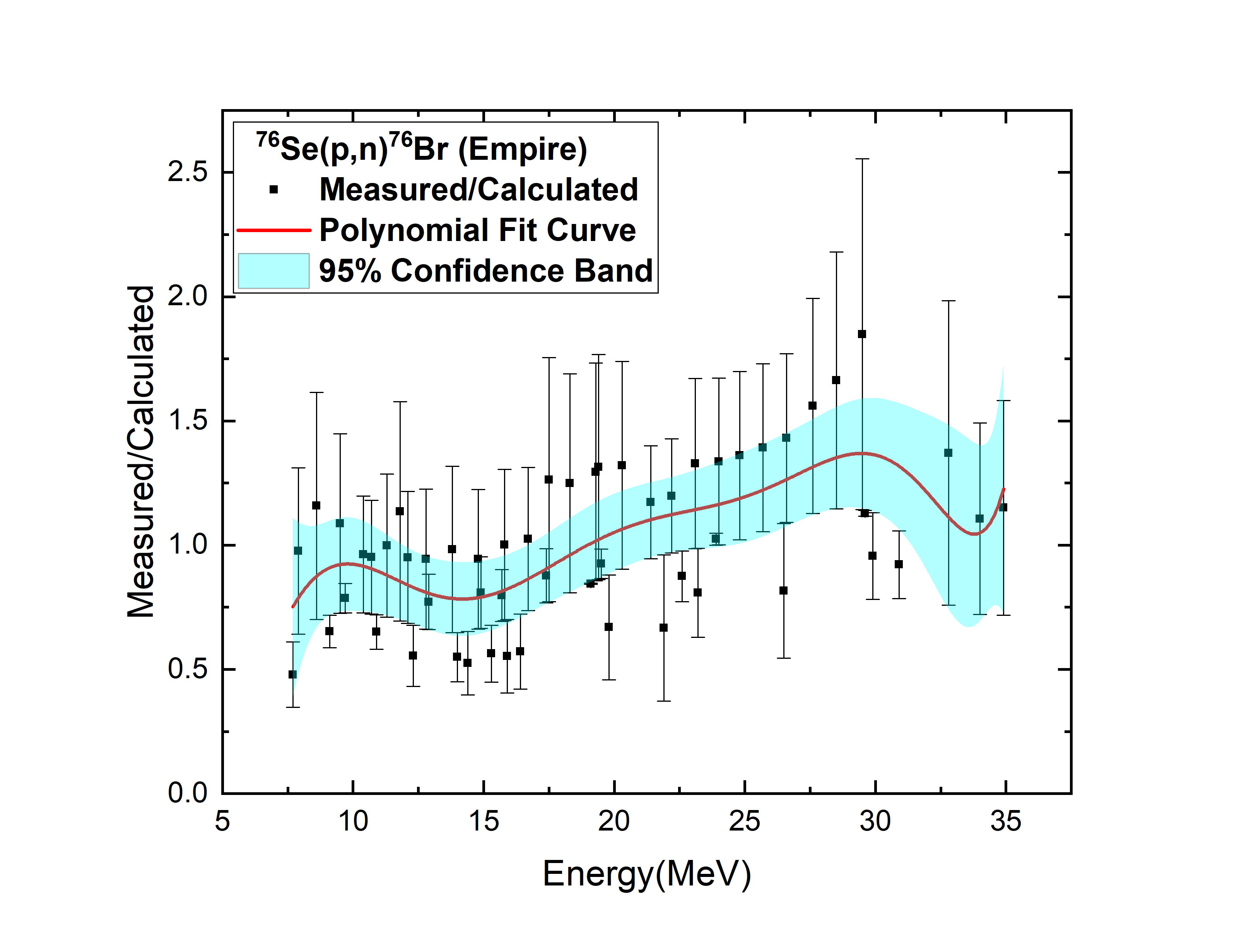}} \\
\hspace{1pt}%
\subfigure[][]{%
\label{fig:ex3-d}%
\includegraphics[height=2.3in]{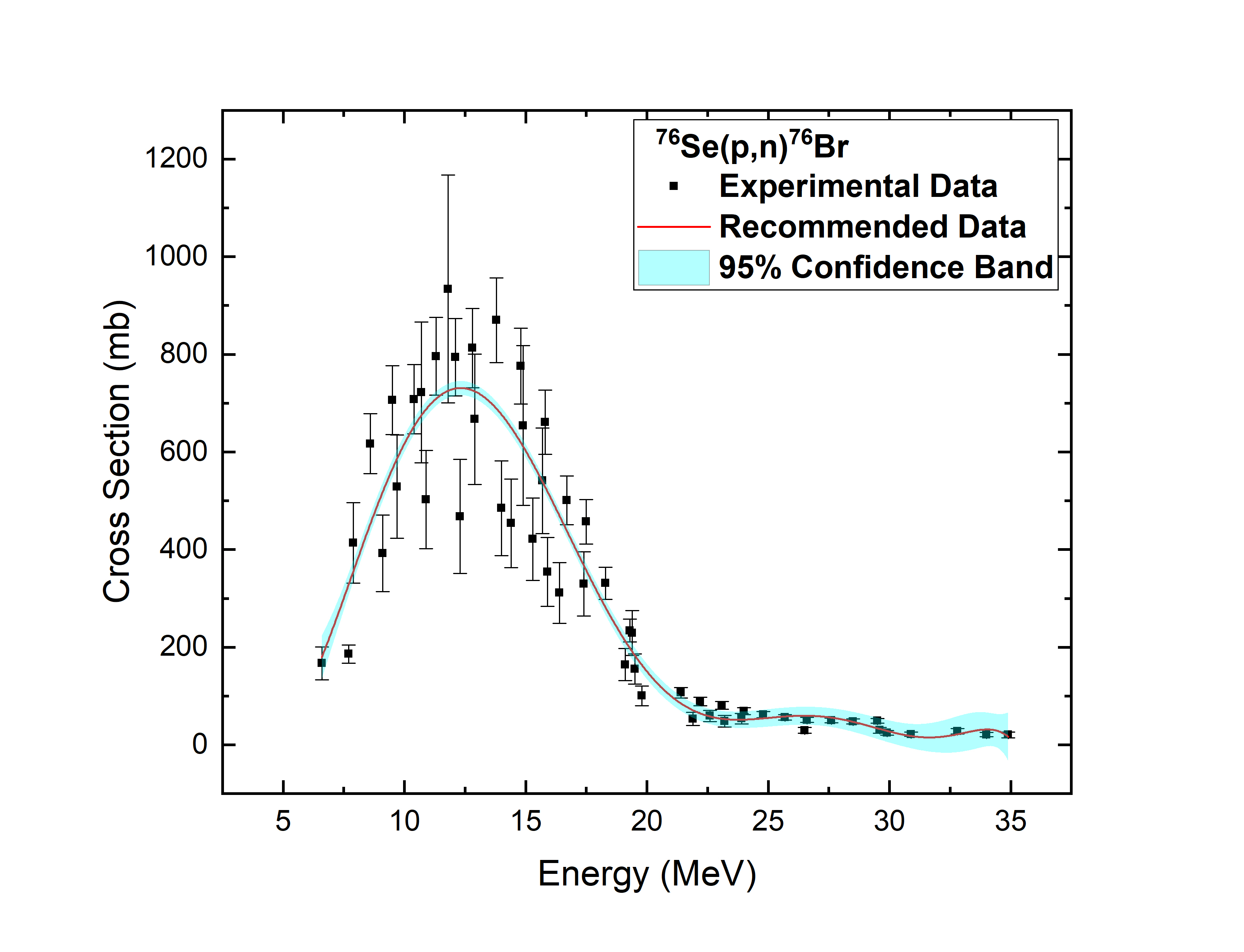}}%
\caption[A set of three subfigures.]{A set of three subfigures:
\subref{fig:ex3-a} Measured by calculated data for $^{76}$Se(p,n)$^{76}$Br reaction by TALYS-1.95 code;
\subref{fig:ex3-b} Measured by calculated data for $^{76}$Se(p,n)$^{76}$Br reaction by EMPIRE-3.1.1 code; and 
\subref{fig:ex3-d} Comparison of our recommended cross sections with experimental data for $^{76}$Se(p,n)$^{76}$Br reaction.}%
\end{center}
\end{figure*}
%%%%%%%%%%%%%%%%%%%%%%%%%%%%%%%%%%%%%%%%%%%%%%%%%%%%%%%%%%%%%%%%%%%%%%%%%%%%%%%%%%%%%%%%%%%%%%%%%%%%%%%%%%%%%%%%%%%%%%%%%%%%%%%%%%%%%%%%%%%%%%%%%%%%%%%%%%%%%%%%%%
\begin{figure*}%
\begin{center}
\subfigure[][]{%
\label{fig:ex3-a}%
\includegraphics[height=2.3in]{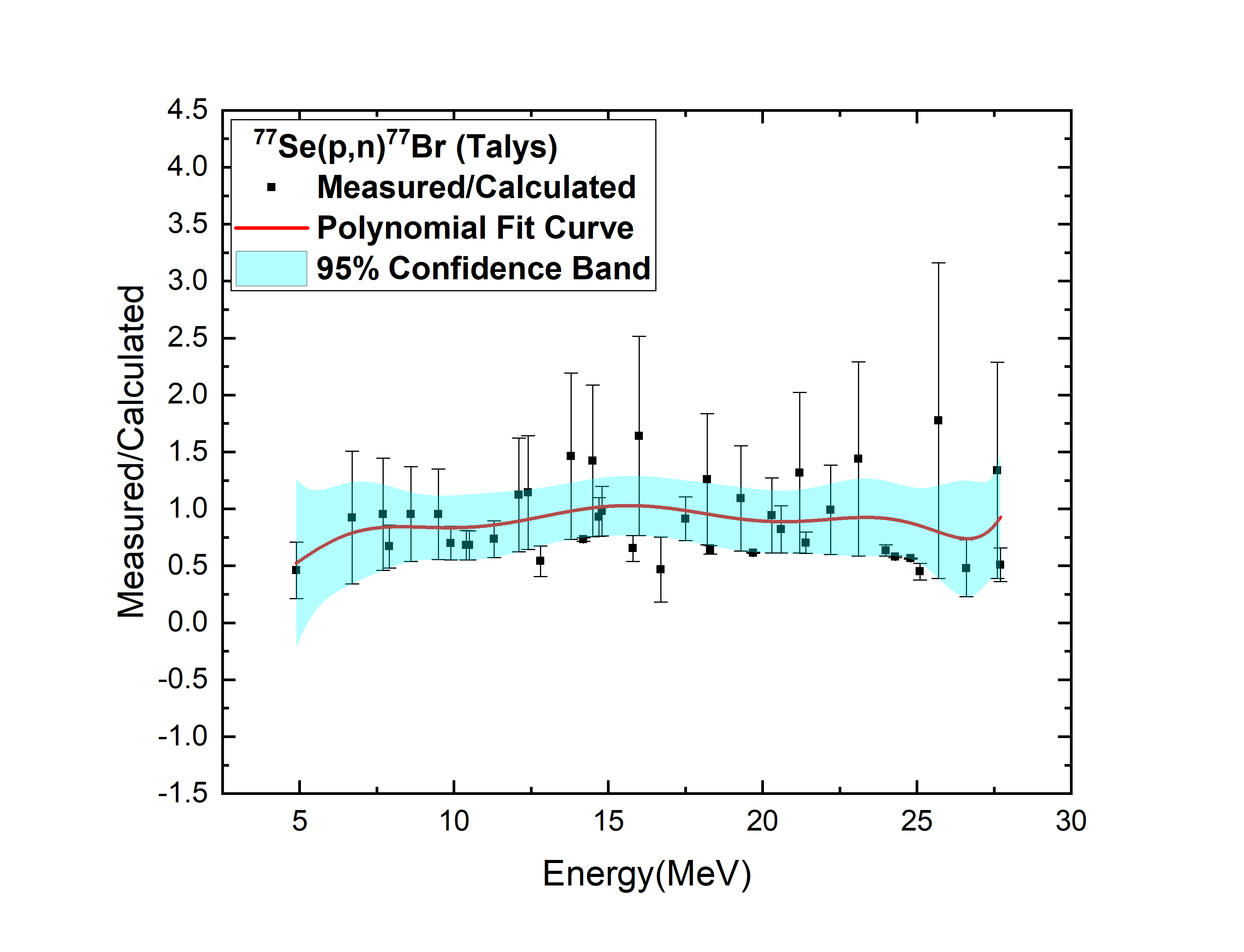}}%
\hspace{1pt}%
\subfigure[][]{%
\label{fig:ex3-b}%
\includegraphics[height=2.3in]{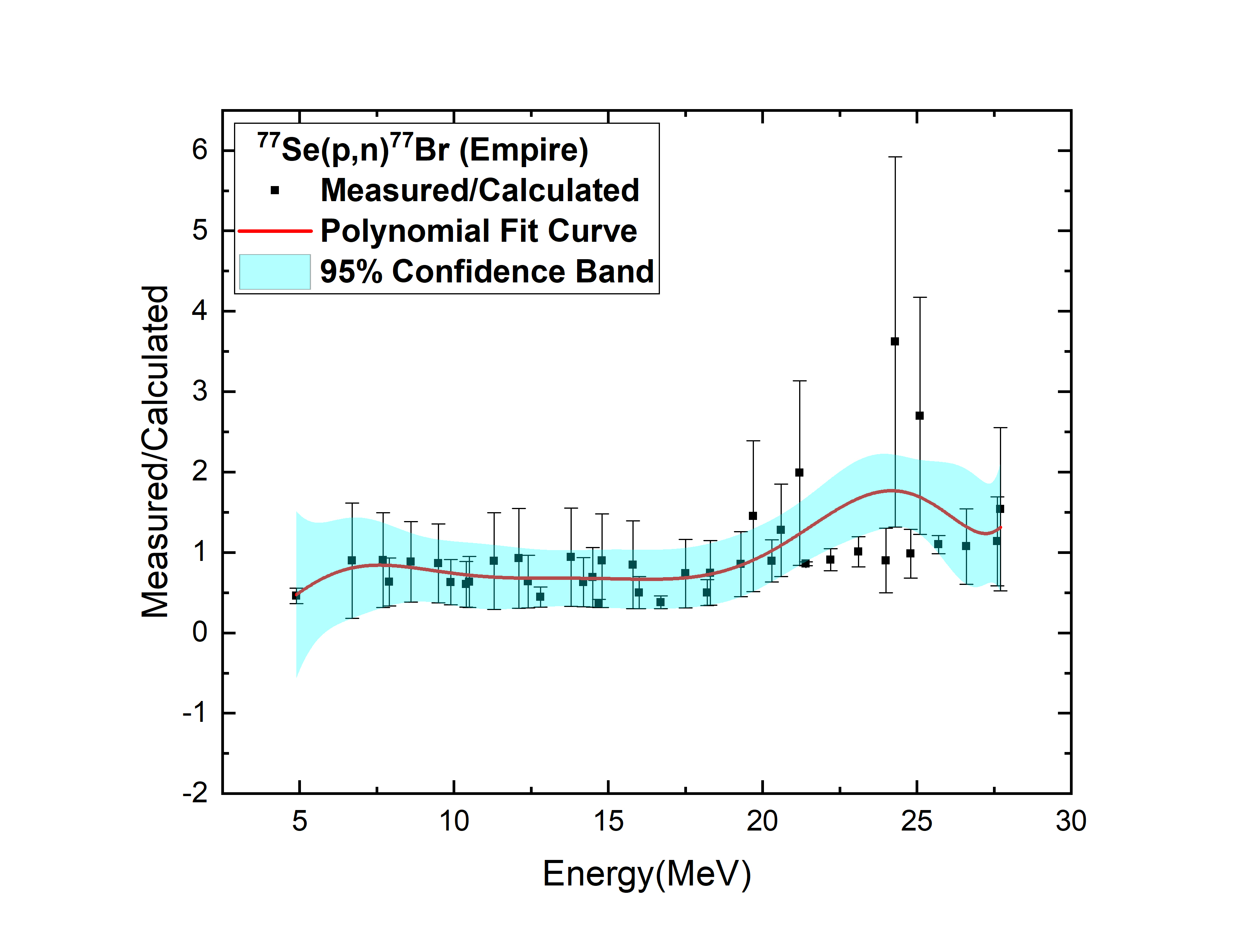}}\\
\hspace{1pt}%
\subfigure[][]{%
\label{fig:ex3-d}%
\includegraphics[height=2.3in]{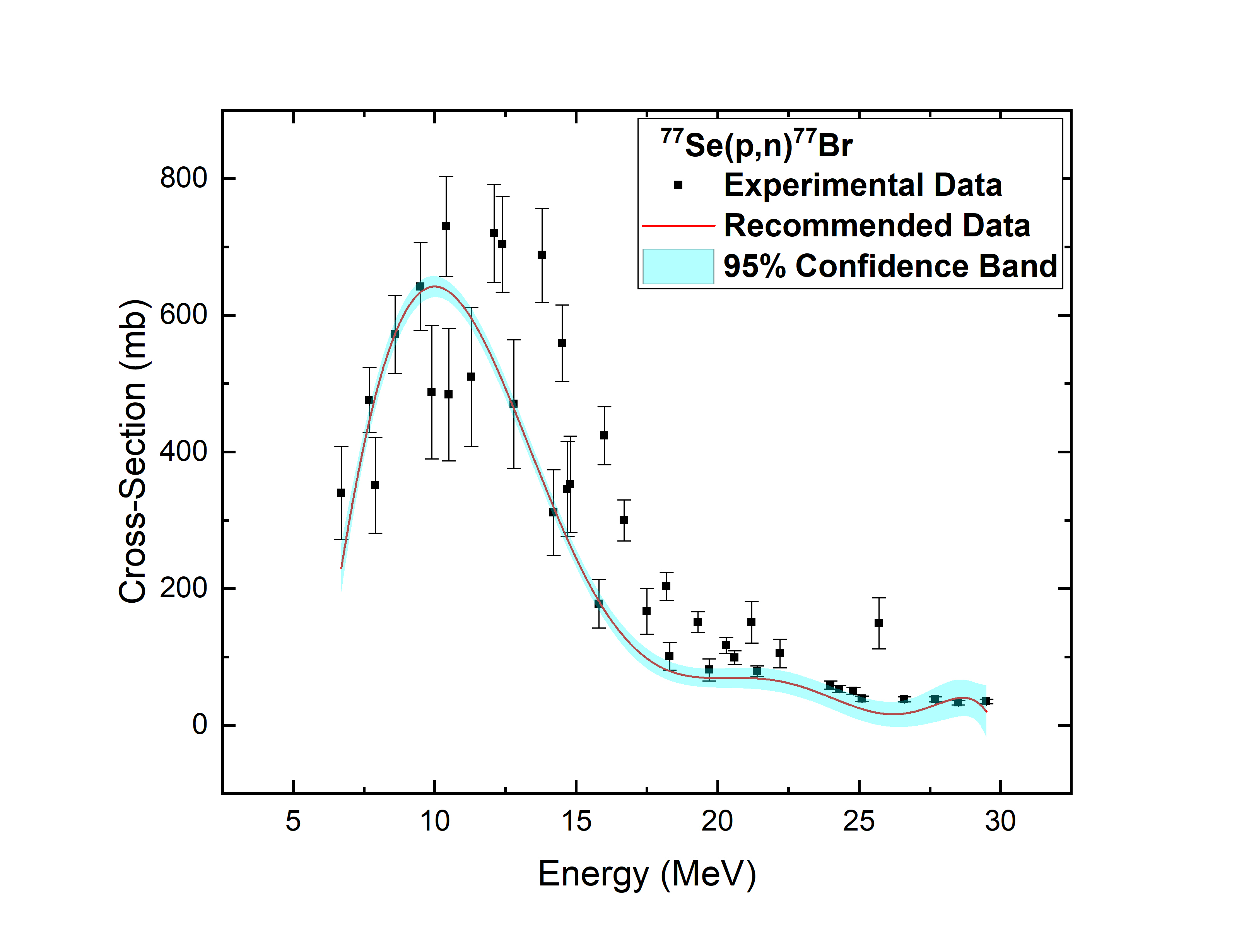}}%
\caption[A set of three subfigures.]{A set of three subfigures:
\subref{fig:ex3-a} Measured by calculated data for $^{77}$Se(p,n)$^{77}$Br reaction by TALYS-1.95 code;
\subref{fig:ex3-b} Measured by calculated data for $^{77}$Se(p,n)$^{77}$Br reaction by EMPIRE-3.1.1 code;
\subref{fig:ex3-d} Comparison of our recommended cross sections with experimental data for $^{77}$Se(p,n)$^{77}$Br reaction.}%
\label{fig:ex3}%
\end{center}
\end{figure*}
%%%%%%%%%%%%%%%%%%%%%%%%%%%%%%%%%%%%%%%%%%%%%%%%%%%%%%%%%%%%%%%%%%%%%%%%%%%%%
\begin{figure*}%
\begin{center}
\subfigure[][]{%
\label{fig:ex3-a}%
\includegraphics[height=2.3in]{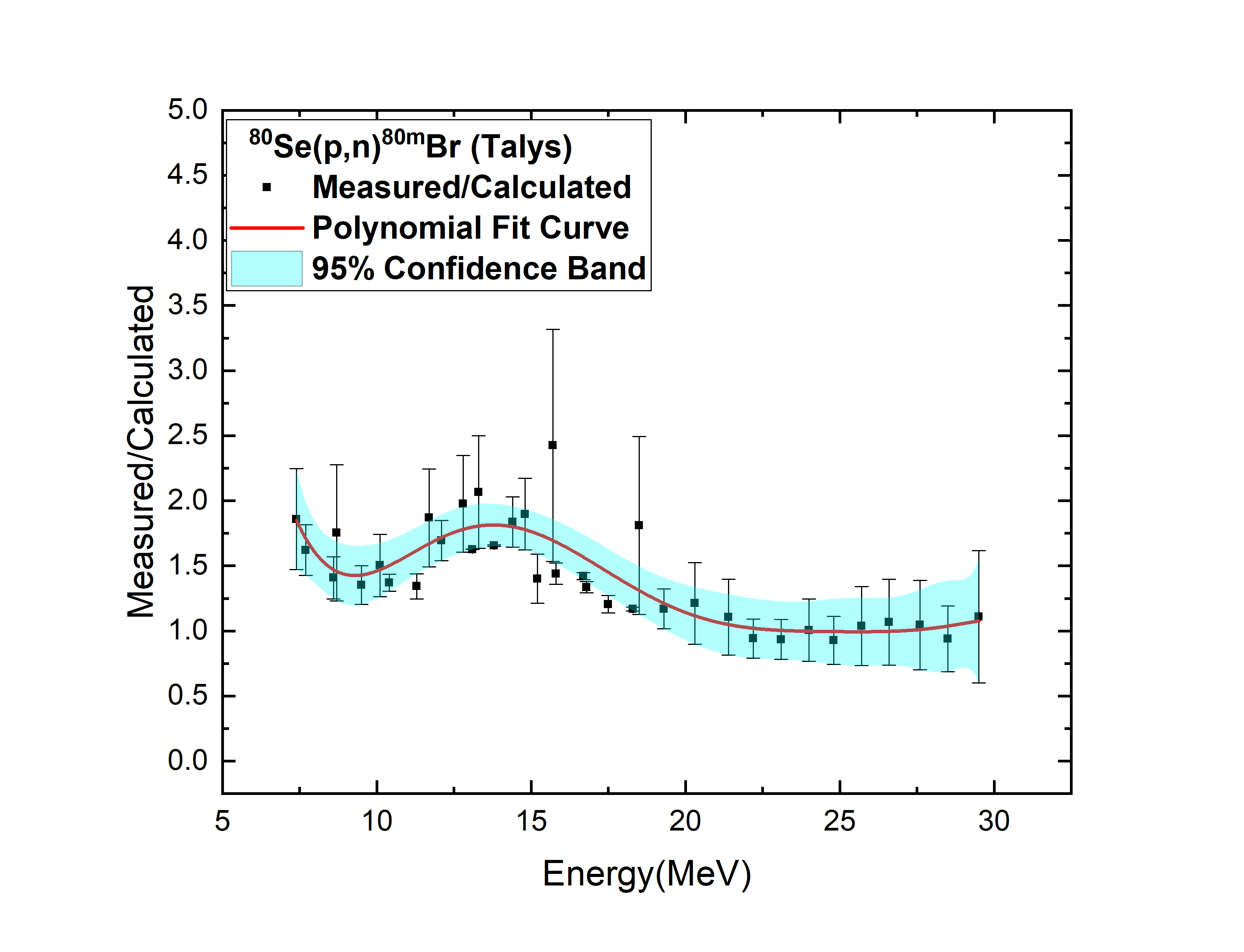}}%
\hspace{1pt}%
\subfigure[][]{%
\label{fig:ex3-b}%
\includegraphics[height=2.3in]{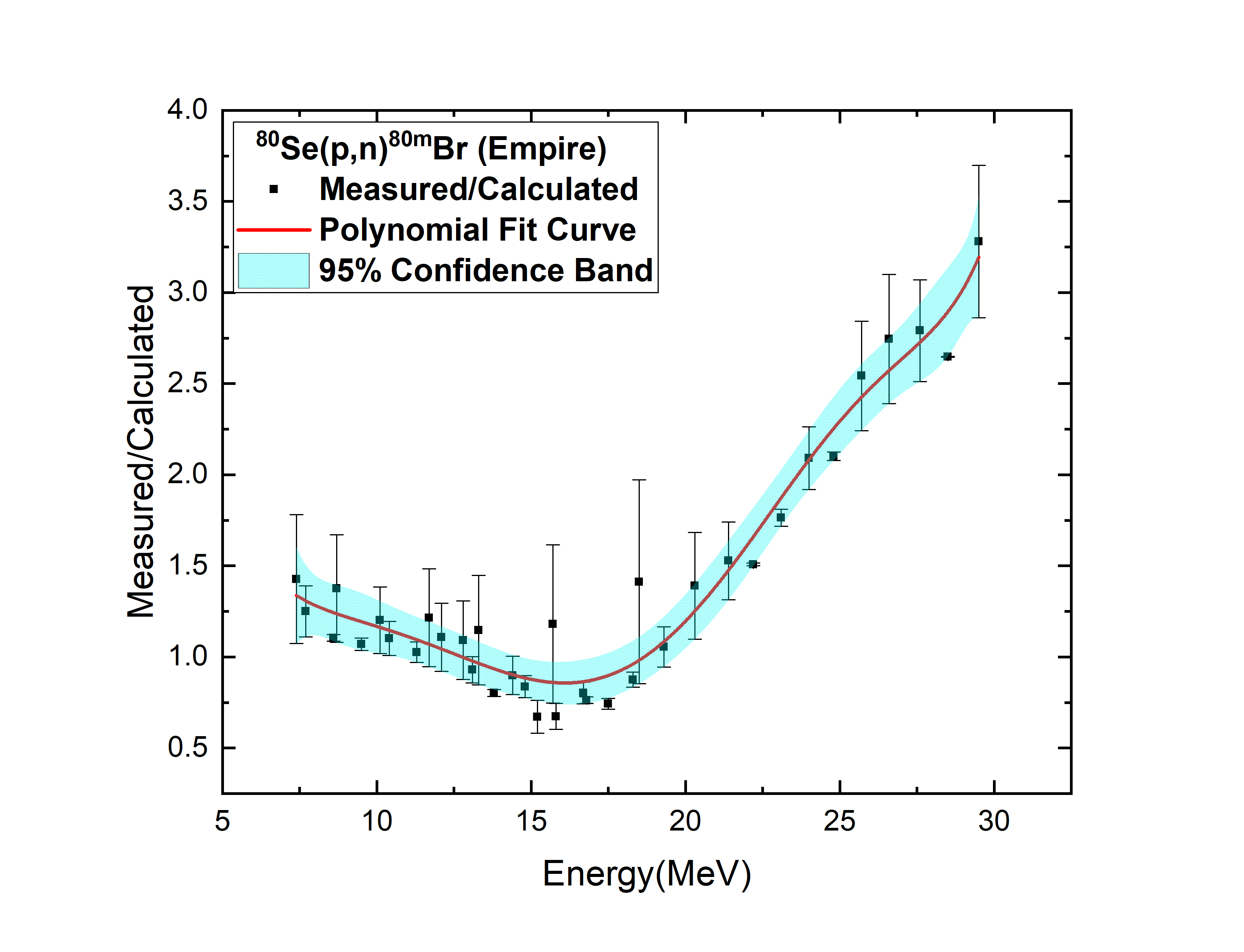}} \\
\hspace{1pt}%
\subfigure[][]{%
\label{fig:ex3-d}%
\includegraphics[height=2.3in]{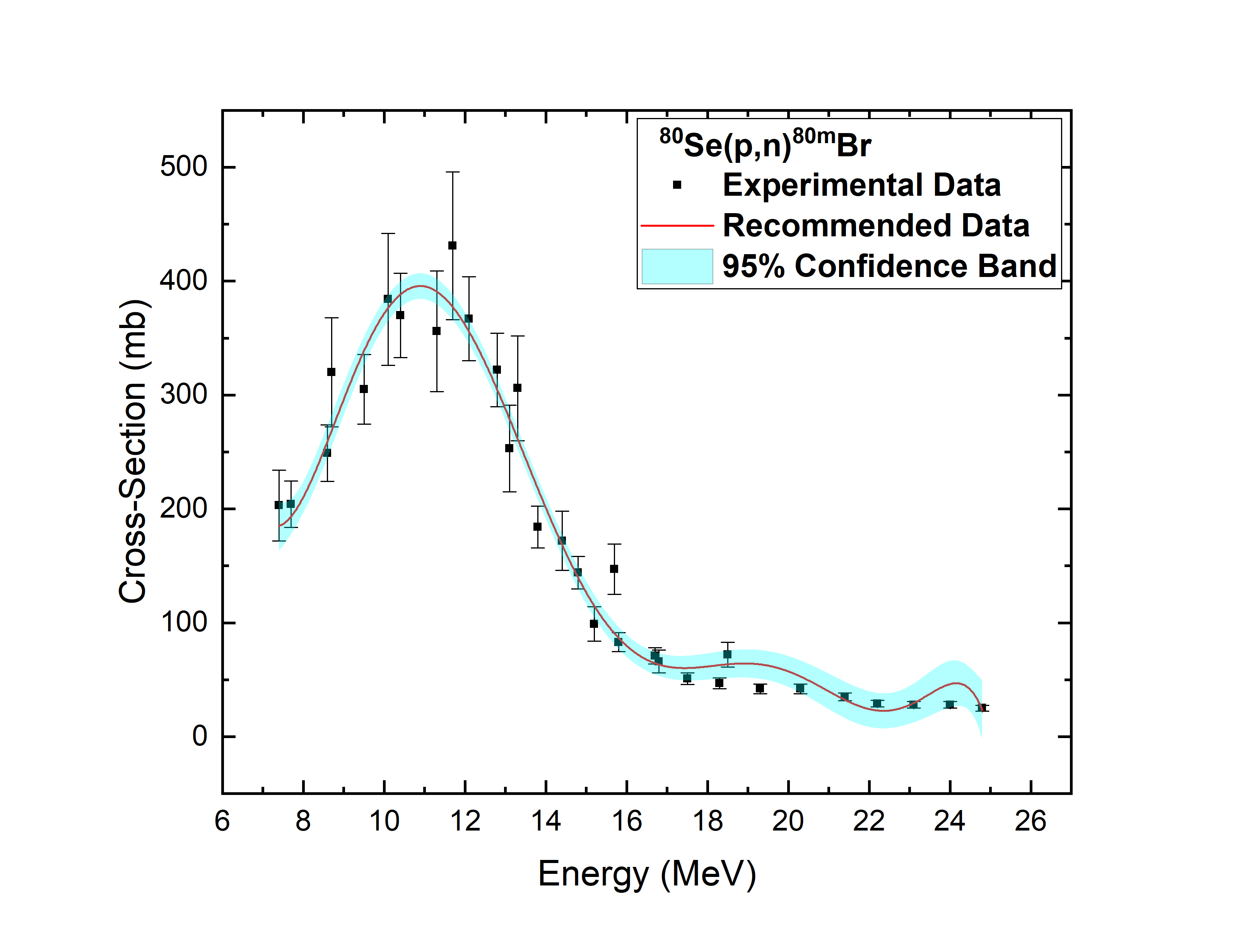}}%
\caption[A set of three subfigures.]{A set of three subfigures:
\subref{fig:ex3-a} Measured by calculated data for $^{80}$Se(p,n)$^{80m}$Br reaction by TALYS-1.95 code;
\subref{fig:ex3-b} Measured by calculated data for $^{80}$Se(p,n)$^{80m}$Br reaction by EMPIRE-3.1.1 code;
\subref{fig:ex3-d} Comparison of our recommended cross sections with experimental data for $^{80}$Se(p,n)$^{80m}$Br reaction.}
\label{fig:ex3}%
\end{center}
\end{figure*}
%%%%%%%%%%%%%%%%%%%%%%%%%%%%%%%%%%%%%%%%%%%%%%%%%%%%%%%%%%%%%%%%%%%%%%%%%%%%
%%%%%%%%%%%%%%%%%%%%%%%%%%%%%%%%%%%%%%%%%%%%%%%%%%%%%%%%%%%%%%%%%%%%%%%%%%%%%%%%%%%%%
\begin{figure*}%
\begin{center}
\subfigure[][]{%
\label{fig:ex3-b}%
\includegraphics[height=2.3in]{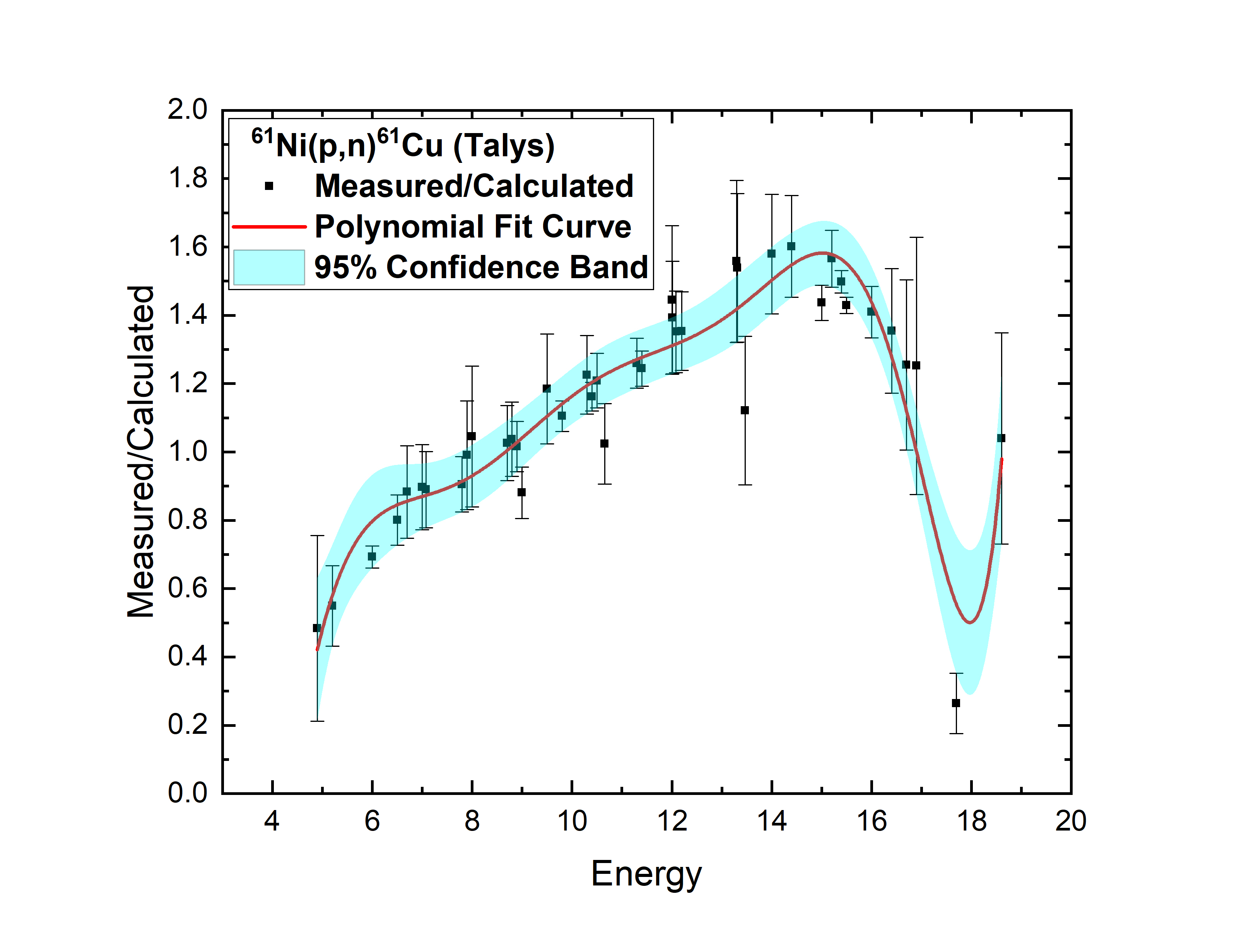}}%
\subfigure[][]{%
\label{fig:ex3-c}%
\includegraphics[height=2.3in]{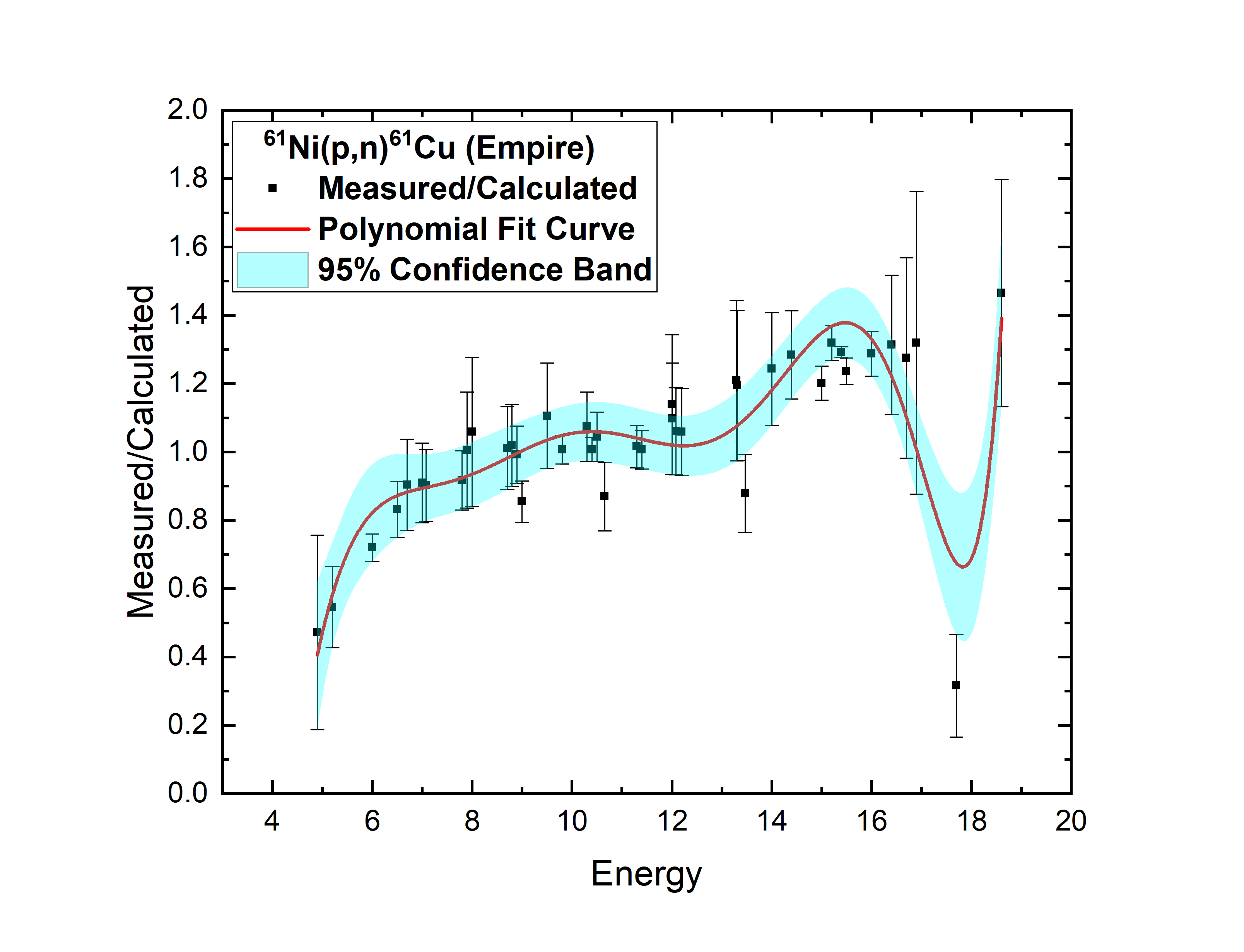}}\\
\hspace{1pt}%
\subfigure[][]{%
\label{fig:ex3-e}%
\includegraphics[height=2.3in]{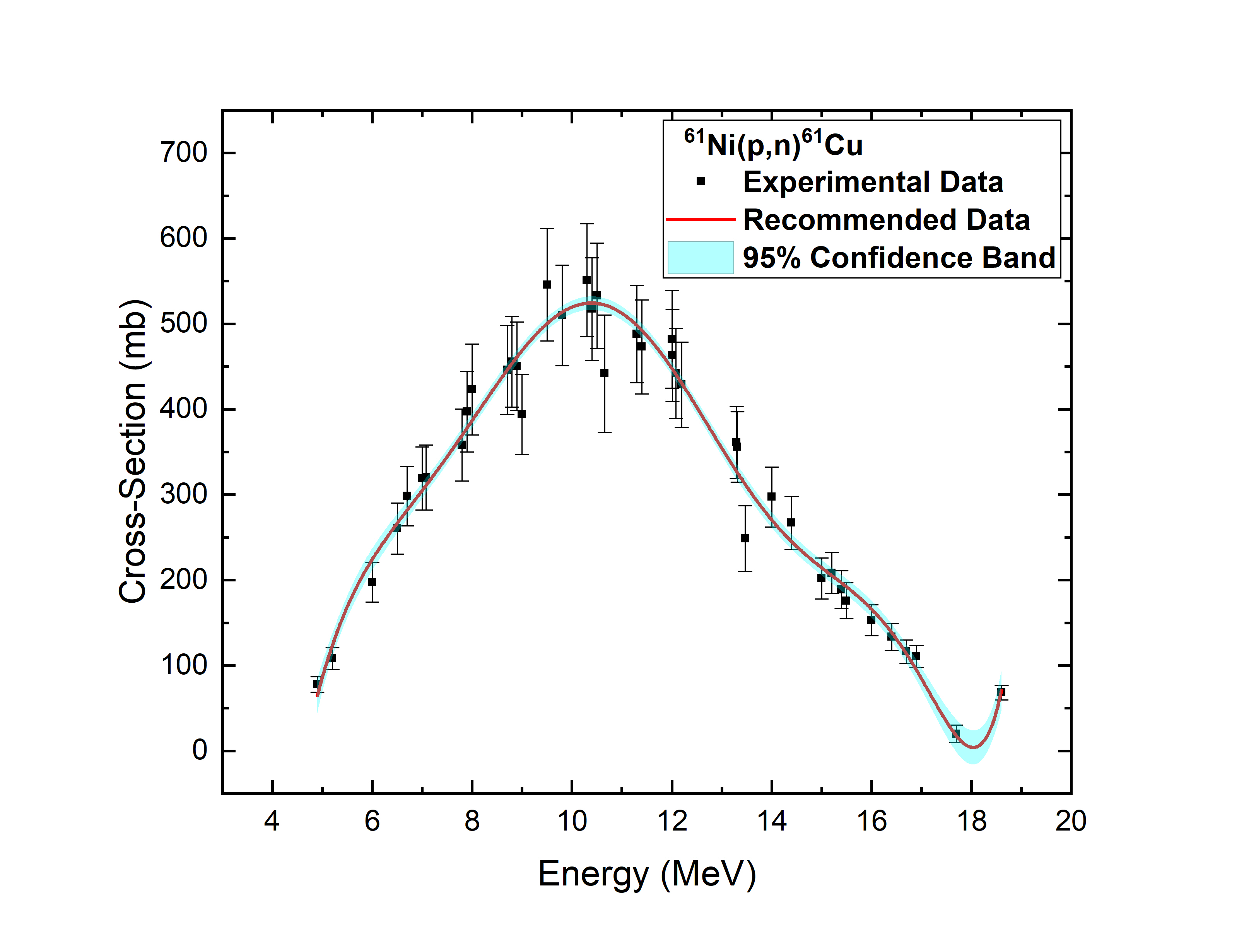}}%
\caption[A set of three subfigures.]{A set of three subfigures:
\subref{fig:ex3-a} Measured by calculated data for $^{61}$Ni(p,n)$^{61}$Cu reaction by TALYS-1.95 code;
\subref{fig:ex3-b} Measured by calculated data for $^{61}$Ni(p,n)$^{61}$Cu reaction by EMPIRE-3.1.1 code;
\subref{fig:ex3-e}  Comparison of our recommended cross sections with experimental data for $^{61}$Ni(p,n)$^{61}$Cu reaction.}%
\label{fig:ex3}%
\end{center}
\end{figure*}
%%%%%%%%%%%%%%%%%%%%%%%%%%%%%%%%%%%%%%%%%%%%%%%%%%%%%%%%%%%%%%%%%%%%%%%%%%%
\begin{figure*}%
\begin{center}
\subfigure[][]{%
\label{fig:ex3-a}%
\includegraphics[height=2.3in]{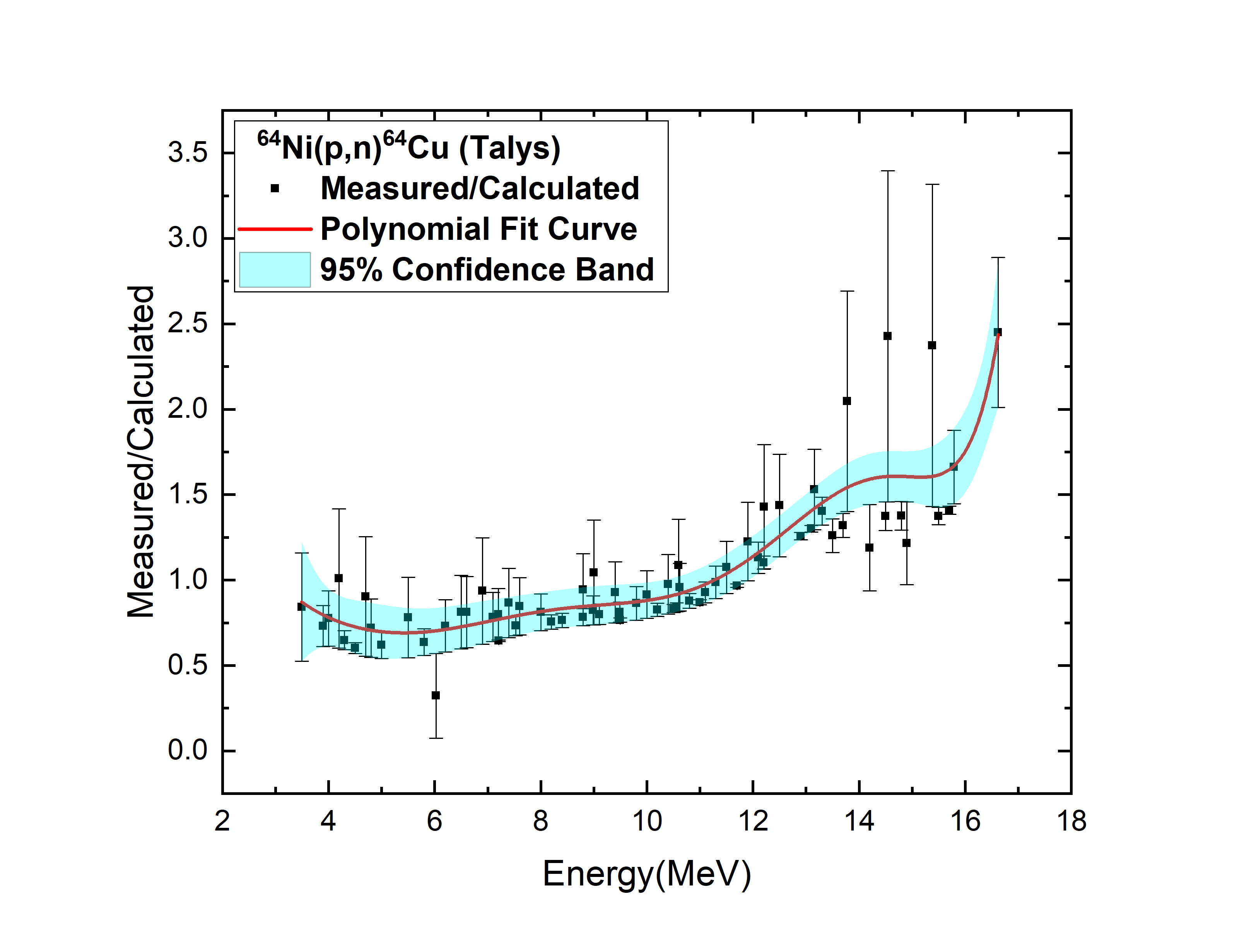}}%
\hspace{1pt}%
\subfigure[][]{%
\label{fig:ex3-b}%
\includegraphics[height=2.3in]{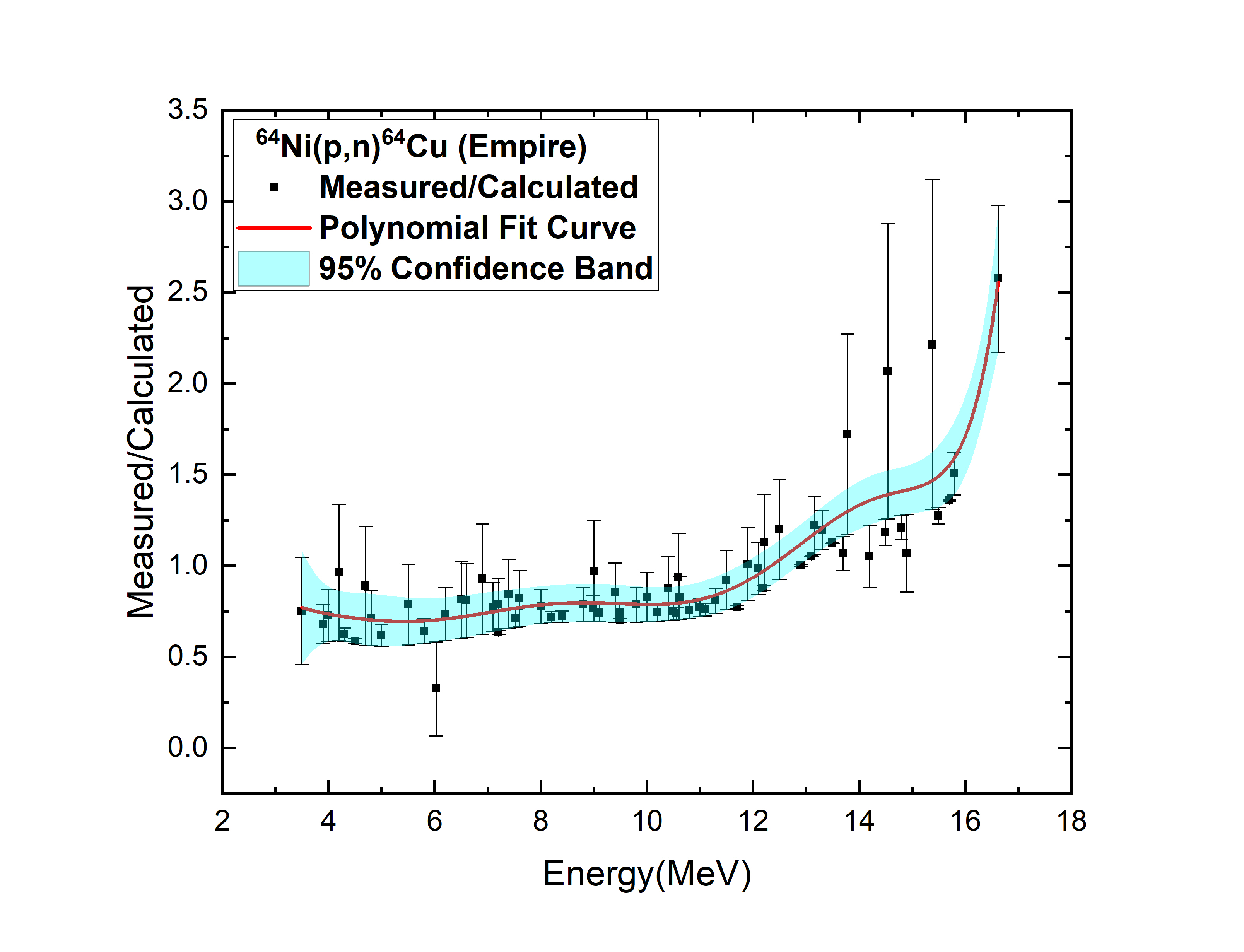}} \\
\hspace{1pt}%
\subfigure[][]{%
\label{fig:ex3-d}%
\includegraphics[height=2.3in]{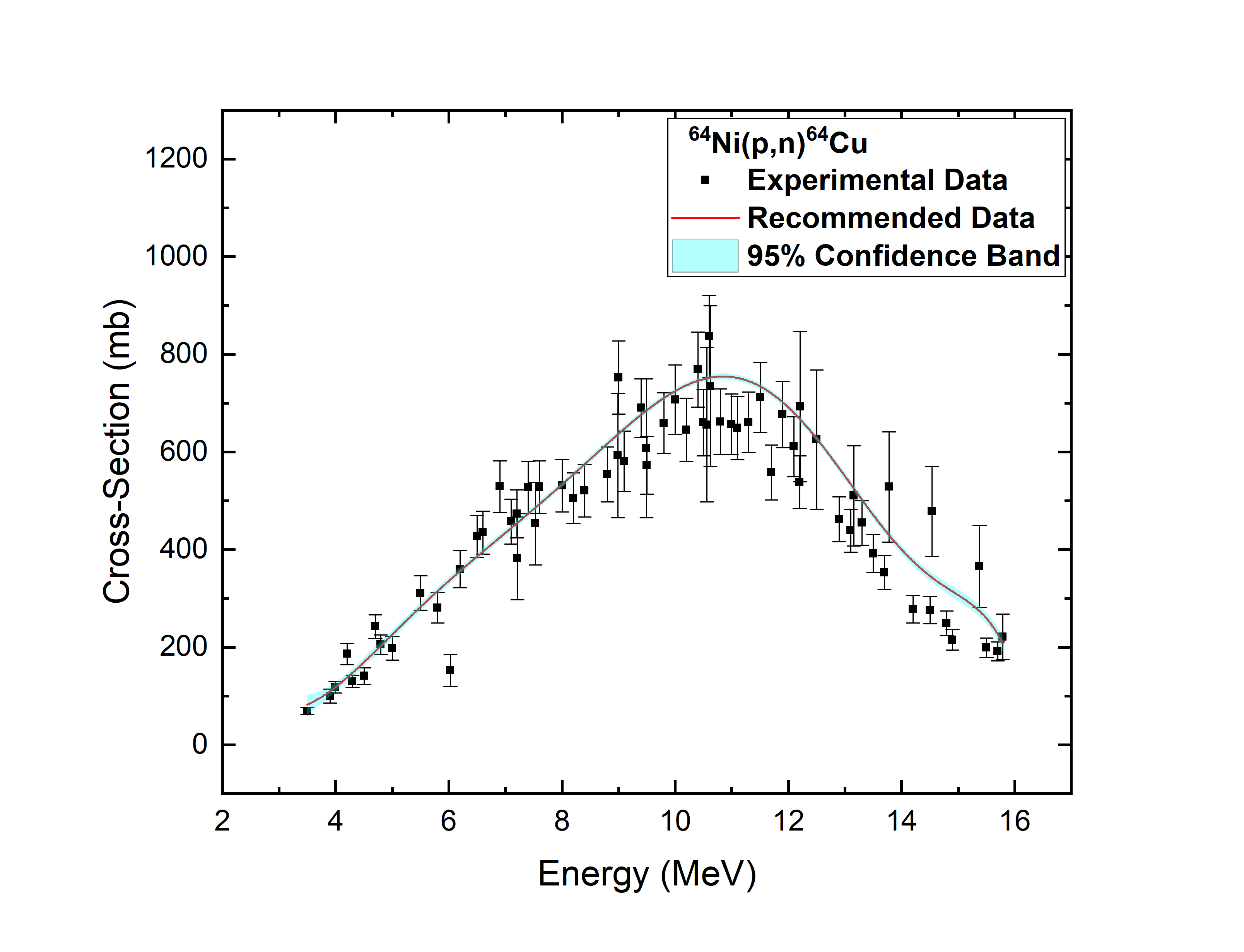}}%
\caption[A set of three subfigures.]{A set of three subfigures:
\subref{fig:ex3-a} Measured by calculated data for $^{64}$Ni(p,n)$^{64}$Cu reaction by TALYS-1.95 code;
\subref{fig:ex3-b} Measured by calculated data for $^{64}$Ni(p,n)$^{64}$Cu reaction by EMPIRE-3.1.1 code;
\subref{fig:ex3-d} Comparison of our recommended cross sections with experimental data for $^{64}$Ni(p,n)$^{64}$Cu reaction.}%
\label{fig:ex3}%
\end{center}
\end{figure*}
%%%%%%%%%%%%%%%%%%%%%%%%%%%%%%%%%%%%%%%%%%%%%%%%%%%%%%%%%%%%%%%%%%%%%%%%%%%%%%%%%%%%%%%
\begin{figure*}%
\begin{center}
\subfigure[][]{%
\label{fig:ex3-a}%
\includegraphics[height=2.3in]{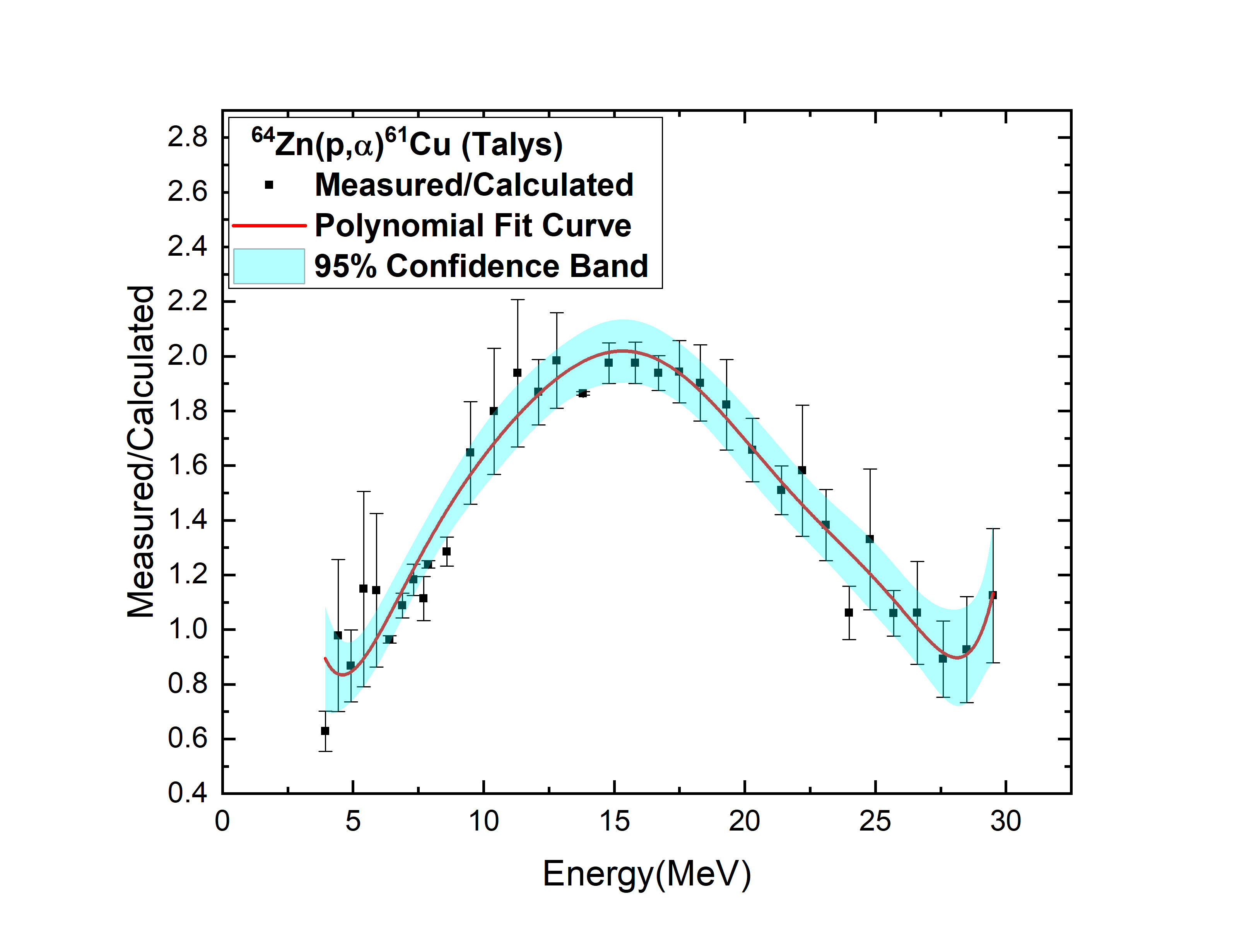}}%
\hspace{1pt}%
\subfigure[][]{%
\label{fig:ex3-b}%
\includegraphics[height=2.3in]{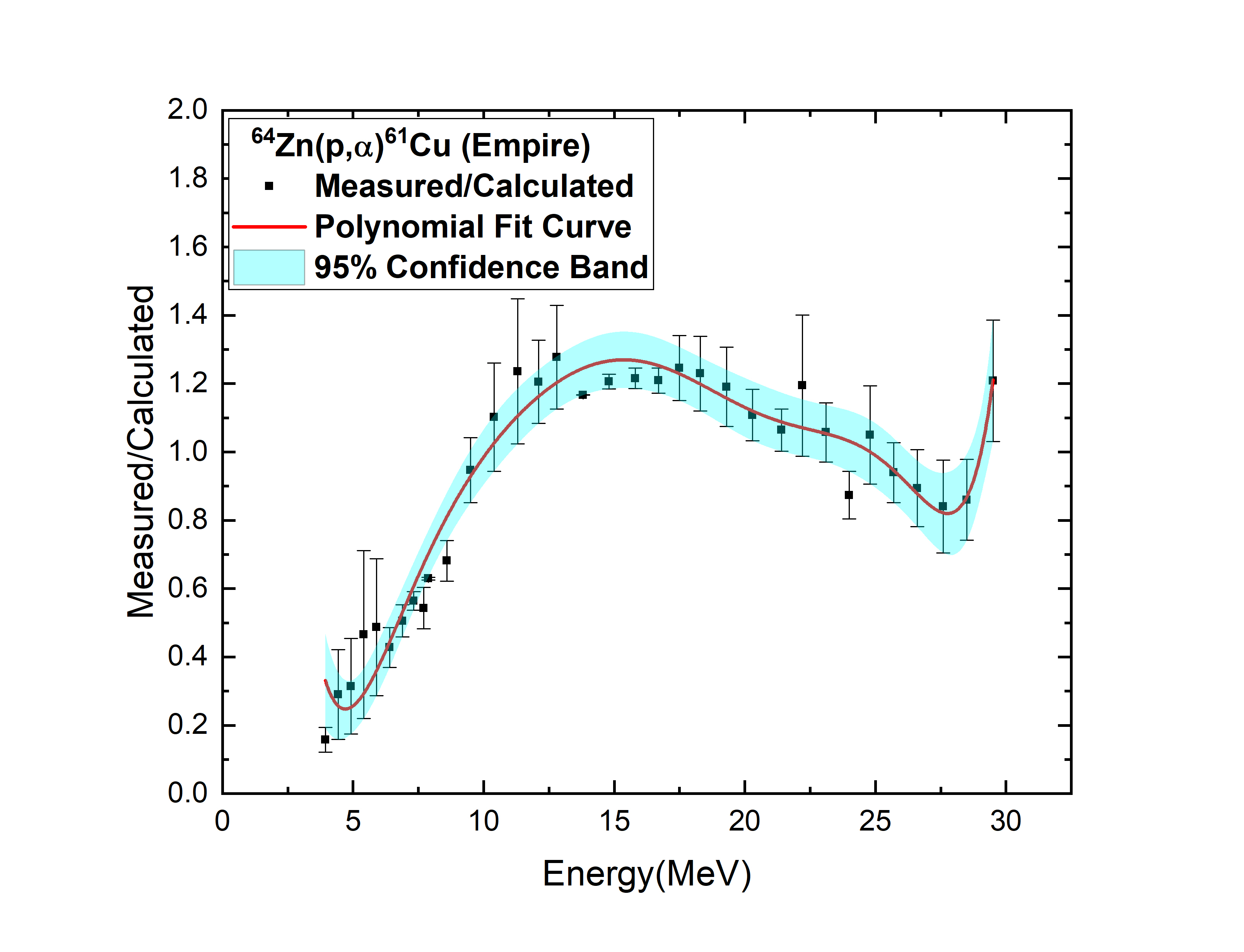}} \\
\hspace{1pt}%
\subfigure[][]{%
\label{fig:ex3-d}%
\includegraphics[height=2.3in]{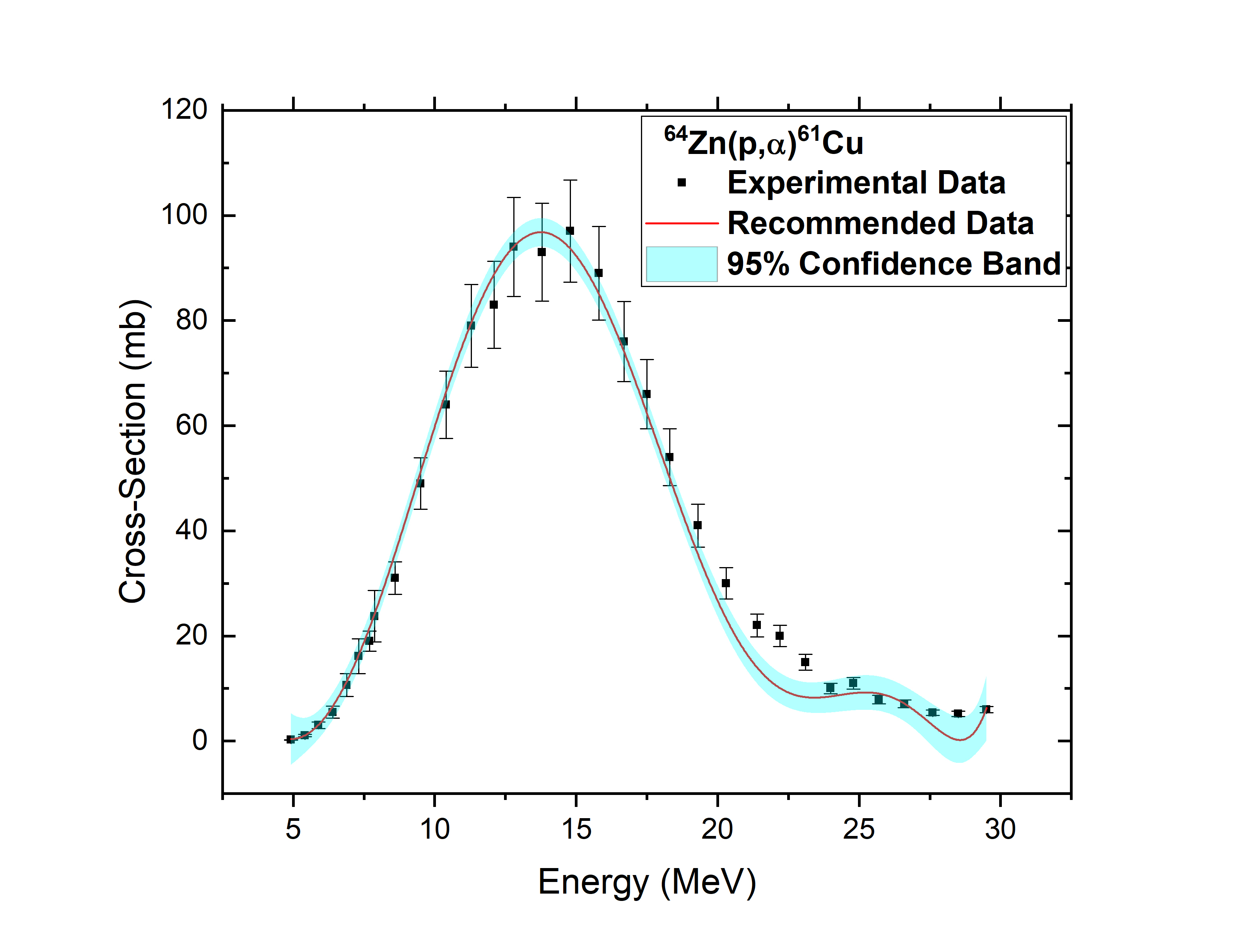}}%
\caption[A set of three subfigures.]{A set of three subfigures:
\subref{fig:ex3-a} Measured by calculated data for $^{64}$Zn(p,$\alpha$)$^{61}$Cu reaction by TALYS-1.95 code;
\subref{fig:ex3-b} Measured by calculated data for $^{64}$Zn(p,$\alpha$)$^{61}$Cu reaction by EMPIRE-3.1.1 code;
\subref{fig:ex3-d} Comparison of our recommended cross sections with experimental data for $^{64}$Zn(p,$\alpha$)$^{61}$Cu reaction.}%
\label{fig:ex3}%
\end{center}
\end{figure*}
%%%%%%%%%%%%%%%%%%%%%%%%%%%%%%%%%%%%%%%%%%%%%%%%%%%%%%%%%%%%%%%%%%%%%%%%%%%%%%%%%%%%%%%%%%%%%%%%%%%%%%%%%%%%%%%%%%%%%%%%%%%%%%%%%%%%%%%%%%%%%%%%%%%%%%%%%%%%%%%%
\begin{figure*}%
\begin{center}
\subfigure[][]{%
\label{fig:ex3-a}%
\includegraphics[height=2.3in]{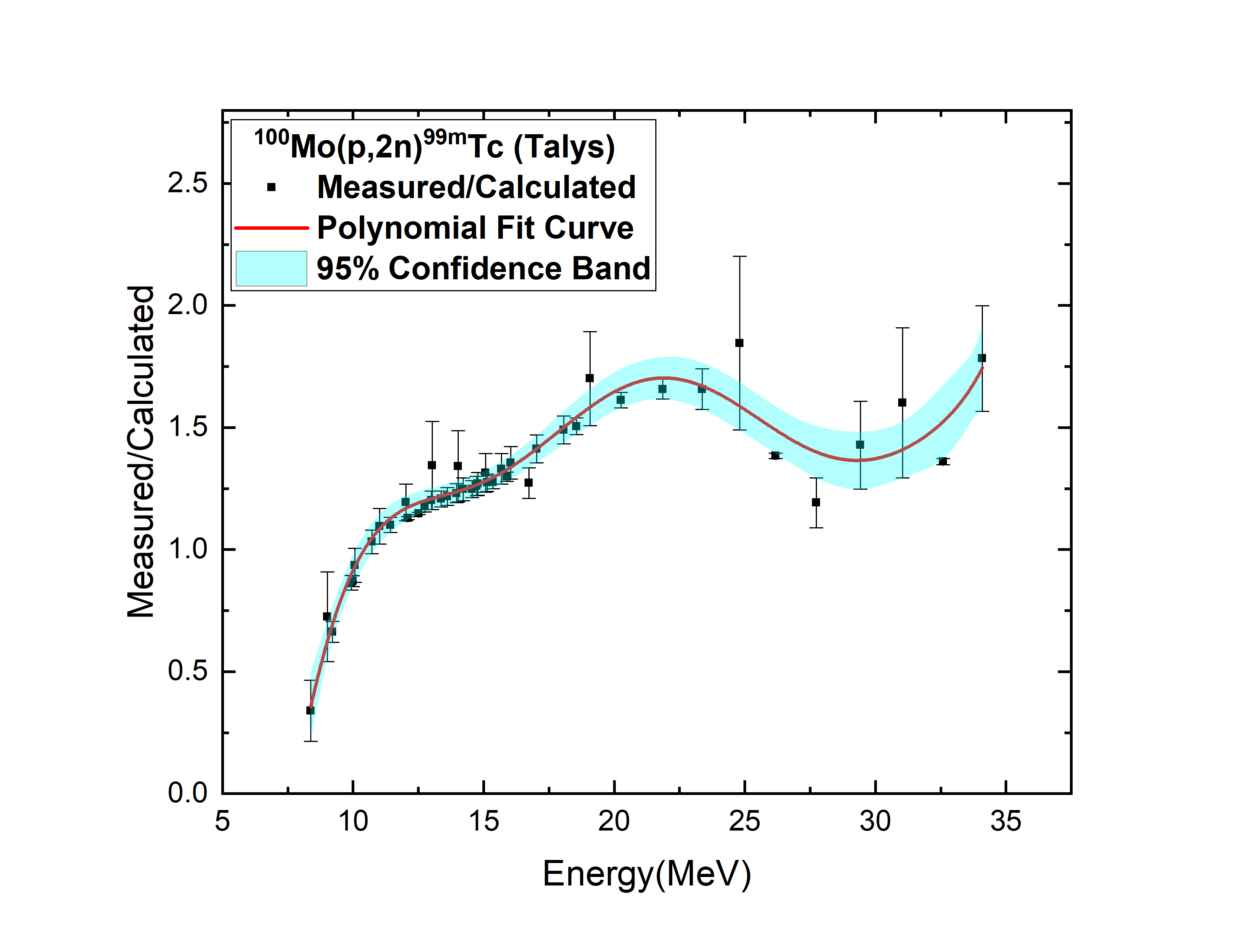}}%
\hspace{1pt}%
\subfigure[][]{%
\label{fig:ex3-b}%
\includegraphics[height=2.3in]{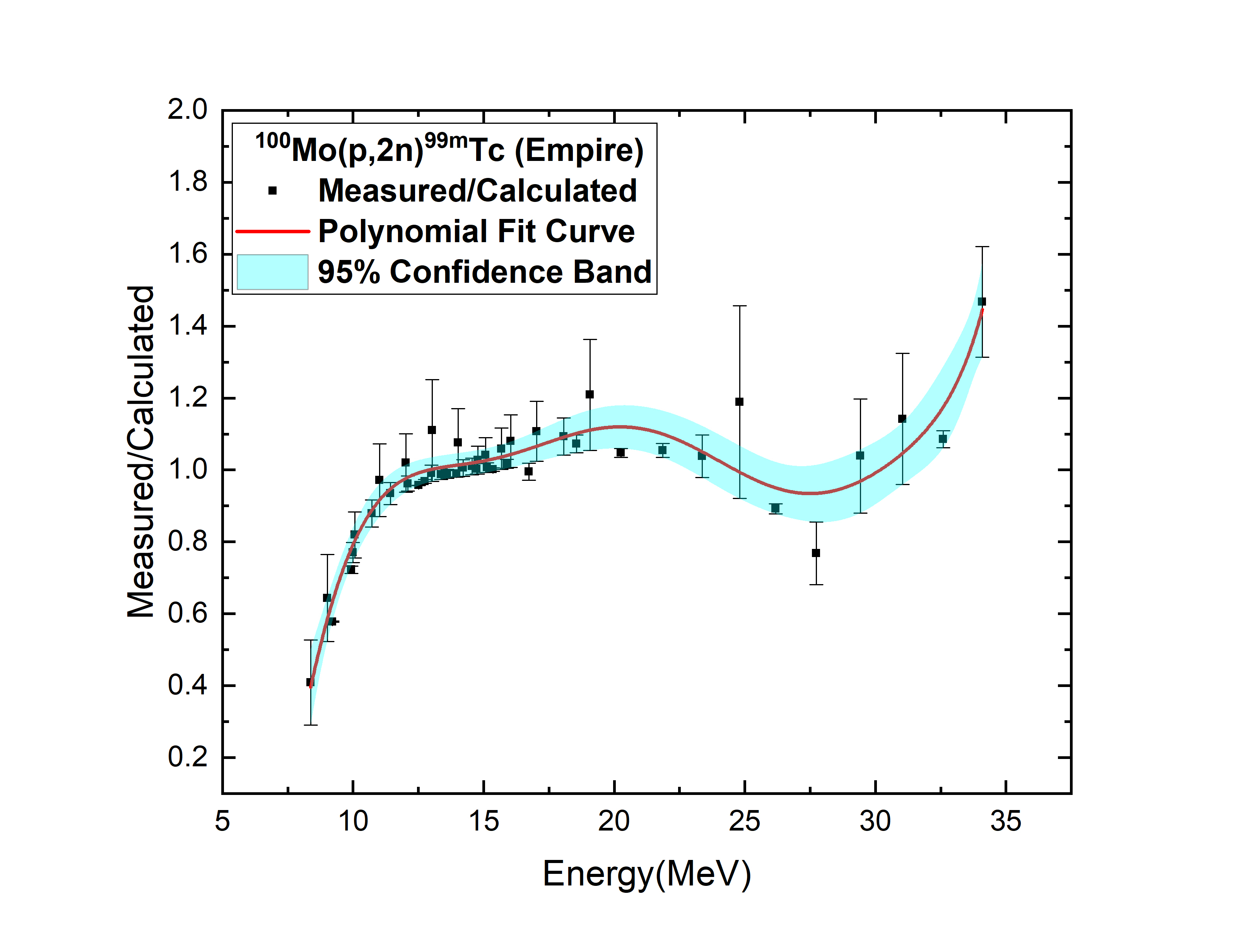}} \\
\hspace{1pt}%
\subfigure[][]{%
\label{fig:ex3-d}%
\includegraphics[height=2.3in]{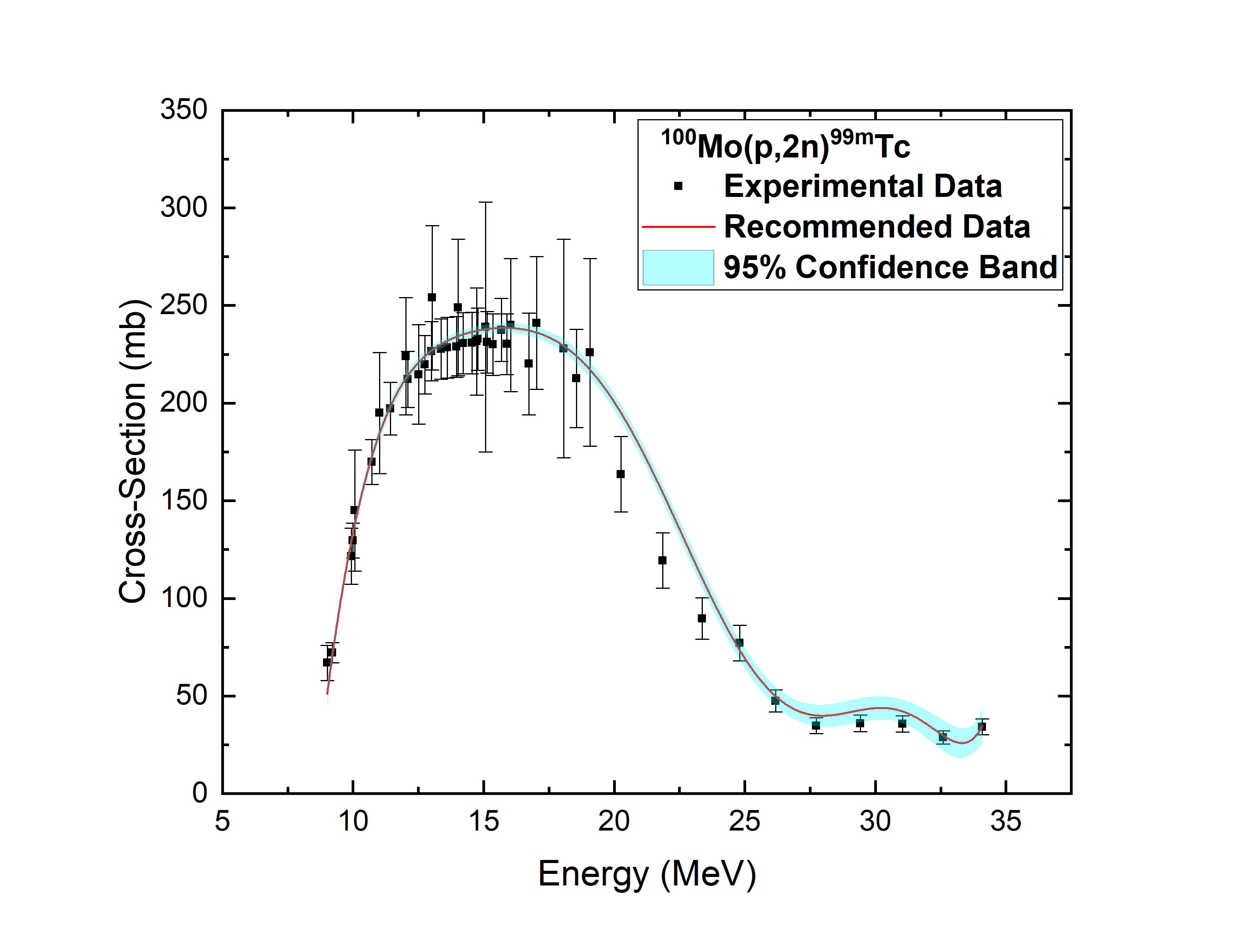}}%
\caption[A set of three subfigures.]{A set of three subfigures:
\subref{fig:ex3-a} Measured by calculated data for $^{100}$Mo(p,2n)$^{99m}$Tc reaction by TALYS-1.95 code;
\subref{fig:ex3-b} Measured by calculated data for $^{100}$Mo(p,2n)$^{99m}$Tc reaction by EMPIRE-3.1.1 code;
\subref{fig:ex3-d} Comparison of our recommended cross sections with experimental data for $^{100}$Mo(p,2n)$^{99m}$Tc reaction.}%
\label{fig:ex3}%
\end{center}
\end{figure*}
%%%%%%%%%%%%%%%%%%%%%%%%%%%%%%%%%%%%%%%%%%%%%%%%%%%%%%%%%%%%%%%%%%%%%%%%%%%%%%%%%%%%%%%%%%%%%%%%%%%%%%%%%%%%%%%%%%%%%%%%%%%%%%%%%%%%%%%%%%%%%%%%%%%%%%%%%%%%%%%%%%
%%%%%%%%%%%%%%%%%%%%%%%%%%%%%%%%%%%%%%%%%%%%%%%%%%%%%%%%%%%%%%%%%%%%%%%%%%%%%%%%%%%%%%
%%%%%%%%%%%%%%%%%%%%%%%%%%%%%%%%%%%%%%%%%%%%%%%%%%%%%%%%%%%%%%%%%%%%%%%%%%%%%%%%%%%%%%%%%%%%%%%%%%%%%%%%
\subsubsection{ $^{76}$Se(p,n)$^{76}$Br reaction}:
Experimental cross sections are obtained from the data reported by V. N. Levkovski~\cite{levkovski1991cross} and Hassan~\cite{hassan2017investigation} and then theoretical model calculations are performed using TALYS-1.95 and EMPIRE-3.1.1. The measured by calculated data are plotted along with incident proton energies to check the deviation of experimental data from theoretical predictions. For all the level density models the measured by calculated data set are calculated and it was found that ldmodel-3 along with PSF-2 produces best result. Related plots are given in Fig 1. 

Recommended cross sections are generated by only considering ratio points lying inside 3$\sigma$  range. For TALYS-1.95 code, ldmodel-3 produces recommended  3$\sigma$ range 0.00-2.12 unit, so for this ldmodel-3, PSF-2 is considered and the range is 0.0-2.10 which is consistent with that of EMPIRE-3.1.1 which is 0.00-2.04. LEVDEN-3 along with GSTRFN-1 yields very good results for EMPIRE-3.1.1. Although some data points are still outside the confidence limit but they are inside the 3$\sigma$ range. As the data distribution is quite scattering in nature we had large $\sigma$ value that is why we have highly fluctuating data points that comes within the 3$\sigma$ range. But they are not within the confidence interval. Deviation from mean is high and including those in confidence interval will cut down precision. In Fig. 1(c), the recommended data is compared with experimental ones for $^{76}$Se(p,n)$^{76}$Br reaction.
%%%%%%%%%%%%%%%%%%%%%%%%%%%%%%%%%%%%%%%%%%%%%%%%%%%%%%%
\subsubsection{ $^{77}$Se(p,n)$^{77}$Br reaction}:
Experimental cross sections are obtained from the data reported by V. N. Levkovski~\cite{levkovski1991cross} and Hassan \cite{hassan2017investigation} and then theoretical simulations are performed using TALYS-1.95 and EMPIRE-3.1.1 codes. The measured by calculated data are plotted along with incident energies to see the deviation of experimental data from theoretical predictions. For all the level density models the measured by calculated data set are calculated and ldmodel-3 along with PSF-2 produces best result and hence this data set is used to predict the recommended cross sections. The obtained $3\sigma$ range is also consistent with that of LEVDEN-3 of EMPIRE-3.1.1 code.

The same definition  of recommended data are also considered here. For TALYS-1.95 code, ldmodel-3 recommended $3\sigma$ range is 0.0-2.5 unit, but for this ldmodel-3, PSF-2 is considered and range is 0.00-1.95 unit which is consistent with that of EMPIRE-3.1.1 where the range is 0.00-1.91. LEVDEN-3 along with GSTRFN-1 yields very good results for EMPIRE-3.1.1. Related plots are given in Fig 2. All the measured by calculated data points are almost inside 95\% confidence limit as well as within the 3$\sigma$ range. In Fig. 2(c), our recommended cross sections are compared with experimental ones for $^{77}$Se(p,n)$^{77}$Br reaction.
%%%%%%%%%%%%%%%%%%%%%%%%%%%%%%%%%%%%%%%%%%%%%%%%%%%%%%%%%%%%%%%%%%%%%%%%%%%%
\subsubsection{$^{80}$Se(p,n)$^{80m}$Br reaction}:
Experimental data are obtained from the data reported by I. Spahn \cite{spahn2010new} and V. N. Levkovski \cite{levkovski1991cross} and then theoretical simulations are performed using TALYS-1.95 and EMPIRE-3.1.1 codes. The measured by calculated data are plotted along with incident proton energies to see the deviation of experimental data from theoretical predictions. For all the level density models the measured by calculated data set are calculated and ldmodel-1 is chosen for TALYS-1.95. For this isomeric reaction recommended range of TALYS-1.95 deviates from that of EMPIRE-3.1.1 code. Recommended range is mentioned in following section.

Recommended data is produced in the same manner as in previous cases. For TALYS-1.95 code, ldmodel-1 recommended $3\sigma$ range as 0.00-2.56 unit which is not consistent with that of EMPIRE-3.1.1 considering LEVDEN-3 with GSTRFN-1. The recommended $3\sigma$ range of EMPIRE-3.1.1 is 0.0-3.3 unit. Although the recommended $3\sigma$ range of TALYS-1.95 and EMPIRE-3.1.1 are not consistent with each other, this did not affect recommended data generation as the ratio points are well inside even for ldmodel-1 of TALYS-1.95. In Fig. 3(c), our recommended cross sections are compared with experimental ones for $^{80}$Se(p,n)$^{80m}$Br reaction.
%%%%%%%%%%%%%%%%%%%%%%%%%%%%%%%%%%%%%%%%%%%%%%%%%%%%%%%%%%%%%%%%%%%%%%%%%%%%
\subsubsection{$^{61}$Ni(p,n)$^{61}$Cu reaction}:
Experimental cross sections are obtained from the data reported by B.P. Singh~\cite{singh2006study} and F. Szelecsenyi \cite{szelecsenyi1993excitation} and then theoretical model calculations are performed using TALYS-1.95 and EMPIRE-3.1.1 codes. The measured by calculated data are plotted along with incident energies to see the deviation of experimental data from theoretical predictions. For all the level density models the measured by calculated data set are calculated and ldmodel-3 of TALYS-1.95 is consistent with that of $3\sigma$ range of EMPIRE-3.1.1.

While  producing the recommended data we find that the cross sections are well equipped inside the 95\% confidence limit here. For TALYS-1.95 code, ldmodel-3 with default PSF produces recommended $3\sigma$ range which is 0.00-2.04 unit, this is consistent with that of EMPIRE-3.1.1 which is 0.00-1.744 unit. For recommended $3\sigma$ range of EMPIRE-3.1.1, LEVDEN-3 with GSTRFN-1 is used. As the data distribution is very smooth in nature, we see deviation or variance is quite less enabling higher precision. Related plots are given in the Fig 4. In Fig. 4(c), our recommended cross sections are compared with experimental ones for $^{61}$Ni(p,n)$^{61}$Cu reaction.
%%%%%%%%%%%%%%%%%%%%%%%%%%%%%%%%%%%%%%%%%%%%%%%%%%%%%%%%%%%%%%%%%%%%%%
\subsubsection{$^{64}$Ni(p,n)$^{64}$Cu reaction}:
Experimental data are obtained from the data by D. Adel \cite{adel2020experimental}, Md.S,Uddin \cite{uddin2016experimental}, R.Adam. Rebeles \cite{rebeles2009new}, M.A. Avila-Rodriguez \cite{avila2007simultaneous} and then theoretical model calculations are performed using TALYS-1.95 and EMPIRE-3.1.1 codes. The measured by calculated data are plotted along with incident proton energies to see the deviation of experimental data from theoretical predictions. For all the level density models the measured by calculated data set are calculated and ldmodel-4 of TALYS-1.95 with default PSF-1 is very much consistent with $3\sigma$ range of EMPIRE-3.1.1 code.

For TALYS-1.95 code, ldmodel-4 with default PSF-1 produces recommended $3\sigma$ range 0.00-2.84 unit, which is in very good agreement with that of EMPIRE-3.1.1 (0.00-2.81) considering LEVDEN-3 along with GSTRFN-1 for EMPIRE-3.1.1. Recommended data is plotted in Fig 7. In Fig. 5(c), our recommended cross section are compared with experimental ones for $^{64}$Ni(p,n)$^{64}$Cu reaction.
%%%%%%%%%%%%%%%%%%%%%%%%%%%%%%%%%%%%%%%%%%%%%%%%%%%%%%%%%%%%%%%
\subsubsection{ $^{64}$Zn(p,$\alpha$)$^{61}$Cu reaction}:
Experimental cross sections are obtained from the data reported by V. N. Levkovski \cite{levkovski1991cross} and Gy. Gyurky \cite{gyurky2014direct} and then theoretical simulations are performed using TALYS-1.95 and EMPIRE-3.1.1 codes. The measured by calculated data are plotted along with incident proton energies to see the deviation of experimental data from theoretical predictions. For all the level density models the measured by calculated data set are calculated and ldmodel-2 along with default PSF matches greatly with $3\sigma$ range of EMPIRE-3.1.1 code.
It is important to use only latest cross sections to avoid  measurement uncertainties. To produce recommended data, some data points had to be excluded as they were lying outside 3$\sigma$  ranges. For TALYS-1.95 code, ldmodel-2 along with PSF-2 recommended $3\sigma$ range is 0.00-1.95 unit which has small deviation with 3$\sigma$ range of EMPIRE-3.1.1 which is 0.00-2.64 unit considering LEVDEN-3 for EMPIRE-3.1.1. In Fig. 6(c), our recommended cross sections compared with experimental ones for $^{64}$Zn(p,$\alpha$)$^{61}$Cu reaction.
%%%%%%%%%%%%%%%%%%%%%%%%%%%%%%%%%%%%%%%%%%%%%%%%%%%%%%%%%%%%%%%%%%%%%%%%%%%%%%%%%
\subsubsection{$^{100}$Mo(p,2n)$^{99m}$Tc reaction}:
Experimental cross sections are obtained from the data reported by E. Lamere \cite{lamere2019proton} and S. Takacs \cite{takacs2015reexamination} and then theoretical simulations are performed using TALYS-1.95 and EMPIRE-3.1.1 codes. The measured by calculated data are plotted along with incident proton energies to see the deviation of experimental data from theoretical predictions. For all the level density models the measured by calculated data set are calculated and ldmodel-4 along with PSF-2 produces best results for TALYS-1.95. Recommended energy range is mentioned in the following section.

 For TALYS-1.95 code, ldmodel-4 produces recommended $3\sigma$ range as 0.00-2.44 unit. For this ldmodel-4, PSF-2 is considered and new $3\sigma$ range is 0.0-2.10 unit which is  consistent with that of EMPIRE-3.1.1 which is 0.00-1.52 unit considering LEVDEN-3 along with GSTRFN-1 for EMPIRE-3.1.1. Ratio points lying outside this range are discarded. Related plots are given in Fig 7. It can be seen that the nature of recommended data is in good agreement with that of the existing experimental data. In Fig. 7(c), our recommended cross sections are compared with the existing experimental data for $^{100}$Mo(p,2n)$^{99m}$Tc reaction.
%%%%%%%%%%%%%%%%%%%%%%%%%%%%%%%%%%%%%%%%%%%%%%%%%%%%%%%%%%%%%%%%%%%%%%
%%%%%%%%%%%%%%%%%%%%%%%%%%%%%%%%%%%%%%%%%%%%%%%%%%%%%%%%%%%%%%%%%%%%%%%
Our recommend cross sections are tabulated in Table 3 and Table 4. 
%%%%%%%%%%%%%%%%%%%%%%%%%%%%%%%%%%%%%%%%%%%%%%%%%%%%%%%%%%%%%%%%%%%%%%%%%%%%%%%%%%%%%%%%%%%5
\begin{table*}%[h!]
  \centering
\caption{Recommended data values for the reactions.}
\begin{tabular}{|c|c|c|c|c|c|}
\hline
\multicolumn{6}{c}{Recommended Cross Sections}\\
\hline
\multicolumn{2}{c}{$^{76}$Se(p,n)$^{76}$Br}&\multicolumn{2}{c}{$^{77}$Se(p,n)$^{77}$Br}&\multicolumn{2}{c}{$^{80}$Se(p,n)$^{80m}$Br}\\
\hline
Energy & CS & Energy & CS & Energy & CS \\
(MeV) & (mb) & (MeV) & (mb) & (MeV) & (mb) \\
\hline
6.5	&175.5628 &6	&42.23624	&6.5&	251.36995\\
7.5	&309.64028&	7	&290.56125	&7.5	&186.51594\\
8.5	&446.24695&	8	&475.75105	&8.5	&247.50328\\
9.5	&566.84145&	9	&586.61525	&9.5	&335.72249\\
10.5&	657.4204&	10&	624.94613&	10.5&	389.08659\\
11.5&	709.58709&	11&	601.89527&	11.5&	384.13485\\
12.5&	720.65004&	12&	534.25335&	12.5&	327.75907\\
13.5&	693.01715&	13&	440.96078&	13.5&	244.12838\\
14.5&	633.12191&	14&	340.10827&	14.5&	161.19019\\
15.5&	550.0889&	15&	246.61847&	15.5&	99.92806\\
16.5&	454.31608&	16&	170.73107&	16.5&	68.35948\\
17.5&	356.12244&	17&	117.34591&	17.5&	61.05893\\
18.5&	264.57967&	18&	86.20999&	18.5&	63.79435\\
19.5&	186.61762&	19&	72.86625&	19.5&	61.66745\\
20.5&	126.46366&	20&	70.21332&	20.5&	47.95085\\
21.5&	85.44664&	21&	70.45761&	21.5&	29.61744\\
22.5&	62.16696&	22&	67.17046&	22.5&	24.36007\\
23.5&	53.00472&	23&	57.09474&	23.5&	42.70184\\
24.5&	52.90854&	24&	41.27743&	24.5&	47.60001\\
\hline
\end{tabular}

\end{table*}
%%%%%%%%%%%%%%%%%%%%%%%%%%%%%%%%%%%%%%%%%%%%%%%%%%%%%%%%%%%%%%%%%%%%%%%%%%%%%%%%%%%%
\begin{table*}%[h!]
	\centering
\caption{Recommended data values for the reactions.}
\begin{tabular}{|c|c|c|c|c|c|c|c|}
\hline
\multicolumn{8}{c}{Recommended Cross Sections}\\
\hline
\multicolumn{2}{c}{$^{61}$Ni(p,n)$^{61}$Cu}&\multicolumn{2}{c}{$^{64}$Ni(p,n)$^{64}$Cu}&\multicolumn{2}{c}{$^{64}$Zn(p,$\alpha$)$^{61}$Cu}&\multicolumn{2}{c}{$^{100}$Mo(p,2n)$^{99m}$Tc}\\
\hline
Energy&CS&Energy&CS&Energy&CS&Energy&CS\\
(MeV)&(mb)&(MeV)&(mb)&(MeV)&(mb)&(MeV)&(mb)\\
\hline
4&33.6016&4&120.40081&5&0.16056&9&59.1251\\
4.5&64.83068&4.5&169.50724&6&3.4358&10&134.8840\\
5&97.2422&5&224.858&7&12.56982&11&182.479\\
5.5&133.434&5.5&280.718&8&26.42644&12&210.504\\
6&174.574&6&334.302&9&43.09758&13&226.102\\
6.5&220.292&6.5&385.282&10&60.31142&14&234.335\\
7&268.854&7&434.862&11&75.8057&15&238.134\\
7.5&317.52&7.5&484.702&12&87.64133&16&238.615\\
8&363.002&8&535.919&13&94.43801&17&235.63\\
8.5&401.947&8.5&588.316&14&95.51994&18&228.39\\
9&431.397&9&639.973&15&90.96608&19&216.075\\
9.5&449.167&9.5&687.232&16&81.566&20&198.321\\
10&454.114&10&725.124&17&68.68861&21&175.539\\
10.5&446.269&10.5&748.155&18&54.07756&22&149.037\\
11&426.829&11&751.378&19&39.59368&23&120.918\\
11.5&397.999&11.5&731.596&20&26.93088&24&93.7877\\
12&362.711&12&688.482&21&17.33884&25&70.2988\\
12.5&324.238&12.5&625.379&22&11.39181&26&52.6151\\
13&285.738&13&549.455&23&8.84962&27&41.8738\\
13.5&249.788&13.5&470.858&24&8.66311&28&37.7717\\
14&217.962&14&400.463&25&9.18301&29&38.4154\\
14.5&190.531&14.5&345.725&26&8.63715&30&40.6064\\
15&166.371&15&304.144&27&5.9481&31&40.753\\
15.5&143.173&15.5&253.748&28&1.96892&32&36.6293\\

\hline
\end{tabular}

\end{table*}

\newpage
\section{Conclusion}
The excitation functions of the proton induced reactions on natural Se, Ni, Zn and Mo are studied using TALYS-1.95 and EMPIRE-3.1.1 nuclear reaction model codes, focusing mainly on the production of medical isotopes. Our recommended cross sections for the $^{76}$Se(p,n)$^{76}$Br, $^{77}$Se(p,n)$^{77}$Br, $^{80}$Se(p,n)$^{80m}$Br, $^{61}$Ni(p,n)$^{61}$Cu, $^{64}$Ni(p,n)$^{64}$Cu, $^{64}$Zn(p,$\alpha$)$^{61}$Cu, $^{100}$Mo(p,2n)$^{99m}$Tc reactions are generated using the data available in the EXFOR database and TALYS-1.95 and EMPIRE-3.1.1. For the generation of recommended cross sections normalisation of the reported cross sections are performed using latest decay data and monitor cross sections, wherever possible.

The data analysis is carried out using the concept of data filtration beyond 3$\sigma$ ranges from different data sets. Using the two nuclear reaction model codes, we have fitted polynomial curve from the measured by calculated data at first and 95\% confidence limit is produced for all the plots making sure that this consists maximum r-squared value. For all the level density models of TALYS-1.95, 3$\sigma$  ranges have been calculated and one particular level density model along with one PSF is selected so that the best $3\sigma$ range is obtained. For most of the cases EMPIRE-3.1.1 produces good results as the recommended $3\sigma$ ranges are shorter in nature considering LEVDEN-3 and GSTRFN-1. Finally the best combinations are obtained so that the $3\sigma$ range of TALYS-1.95 and EMPIRE-3.1.1 agrees well.
%So recommended $3\sigma$ ranges for it is quite good because if the ranges becomes higher then we have to include all those data points also which means for those data points measured value is more than theoretically estimated values indicating a sacrifice in precision. So, we those as reference and have chosen a particular ldmodel so that it almost matches the 3$\sigma$ EMPIRE-3.1.1 code. Further problem is faced when ldmodel of TALYS-1.95 alone could not produce appropriate recommended energy ranges. So for that particular chosen ldmodel we varied eight gamma ray strength functions to see the variation of recommended energy ranges.\\
%After calculating the ranges for each cases we see an unique combination of ldmodel and a particular gamma ray strength function recommended energy ranges match finally.

For only $^{80m}$Br production, reduction of recommended energy range could not be possible but fortunately the measured by calculated data lie well within the range. For most of the reactions experimental cross sections lie in the vicinity of theoretically predicted recommended data. 
\section*{Acknowledgments}
One of the authors (R.P.) acknowledged the grants received from the Institutions of Eminence (IoE) BHU (Grant No. 6031-B) and UGC-DAE Consortium for Scientific Research (Grant No. CRS/2021- 22/02/474).
%-------------------------------------- 
{}

\end{document}